\documentclass[pra,superscriptaddress,twocolumn]{revtex4}
\usepackage{graphicx}
\usepackage{bm}
\usepackage{amsmath}
\usepackage{amssymb}
\usepackage{exscale}
\usepackage{wasysym}
\usepackage{ulem}
\usepackage{cancel}
\bibstyle{apsrev.bib}

\usepackage{xcolor}

\begin{document}

\title{Estimating RCP in polydisperse and bidisperse hard spheres via an equilibrium model of crowding}

\author{Carmine Anzivino}

\email[Electronic mail: \ ]{carmine.anzivino@unimi.it,}

\affiliation{Department of Physics ``A. Pontremoli", University of Milan, via Celoria 16, 20133 Milan, Italy}

\author{Mathias Casiulis}

%\email[Electronic mail: \ ]{mc9287@nyu.edu}

\affiliation{Center for Soft Matter Research, Department of Physics, New York University, New York 10003, USA}

\affiliation{Simons Center for Computational Physical Chemistry, Department of Chemistry, New York University, New York 10003, USA}

\author{Tom Zhang}

%\email[Electronic mail: \ ]{mc9287@nyu.edu}

\affiliation{Center for Soft Matter Research, Department of Physics, New York University, New York 10003, USA}

\author{Amgad Salah Moussa}

\affiliation{Syngenta AG, 4058 Basel, Switzerland}

\author{Stefano Martiniani}

%\email[Electronic mail: \ ]{stefano.martiniani@nyu.edu}

\affiliation{Center for Soft Matter Research, Department of Physics, New York University, New York 10003, USA}

\affiliation{Simons Center for Computational Physical Chemistry, Department of Chemistry, New York University, New York 10003, USA}

\affiliation{Courant Institute of Mathematical Sciences, New York University, New York 10003, USA}

\author{Alessio Zaccone}

\email[Electronic mail: \ ]{alessio.zaccone@unimi.it}

\affiliation{Department of Physics ``A. Pontremoli", University of Milan, via Celoria 16, 20133 Milan, Italy}

\date{\today}

\begin{abstract}
We show that an analogy between crowding in fluid and jammed phases of hard spheres captures the density dependence of the kissing number for a family of numerically generated jammed states.
We extend this analogy to jams of mixtures of hard spheres in $d=3$ dimensions, and thus obtain an estimate of the random close packing (RCP) volume fraction, $\phi_{\textrm{RCP}}$, as a function of size polydispersity.
We first consider mixtures of particle sizes with discrete distributions. For binary systems, we show agreement between our predictions and simulations, using both our own and results reported in previous works, as well as agreement with recent experiments from the literature.
We then apply our approach to systems with continuous polydispersity, using three different particle size distributions, namely the log-normal, Gamma, and truncated power-law distributions.
In all cases, we observe agreement between our theoretical findings and numerical results up to rather large polydispersities for all particle size distributions, when using as reference our own simulations and results from the literature.
In particular, we find $\phi_{\textrm{RCP}}$ to increase monotonically with the relative standard deviation, $s_{\sigma}$, of the distribution, and to saturate at a value that always remains below 1.
A perturbative expansion yields a closed-form expression for $\phi_{\textrm{RCP}}$ that quantitatively captures a distribution-independent regime for $s_{\sigma} < 0.5$.
Beyond that regime, we show that the gradual loss in agreement is tied to the growth of the skewness of size distributions.
\end{abstract}

\maketitle

\section{Introduction}

Hard spheres represent one of the most important reference system in statistical mechanics.
This system admits a single control parameter, the fraction of space occupied by the particles, or \textit{volume fraction}, $\phi$, and was initially devised to model the short-range repulsive forces of an idealized atomic liquid.
On the theoretical side, trailblazing simulations by Alder and Wainwright~\cite{Alder1962}, as well as theoretical work by Kirkwood and coworkers~\cite{Kirkwood1933,Kirkwood1942,Boer1949}, led to a wide array of predictions on the behaviour of equilibrium hard spheres that paved the way for models of more complicated liquids.
Pioneering experiments by Pusey, van Megen, Vrij \citep{Pusey_vanMegen,vrij}, since followed by others \citep{Besseling2012}, showed that colloidal systems, such as polymethylmethacrylate (PMMA) and silica particles coated with polymers, can be approximately modelled as hard-sphere fluids.
Overall, the phase behaviour of hard spheres has been studied in great detail and by now it can be said to be well understood \citep{mulero, Hansen_McDonald_BOOK}.

\begin{figure}
    \centering
    \includegraphics[width=0.48\columnwidth]{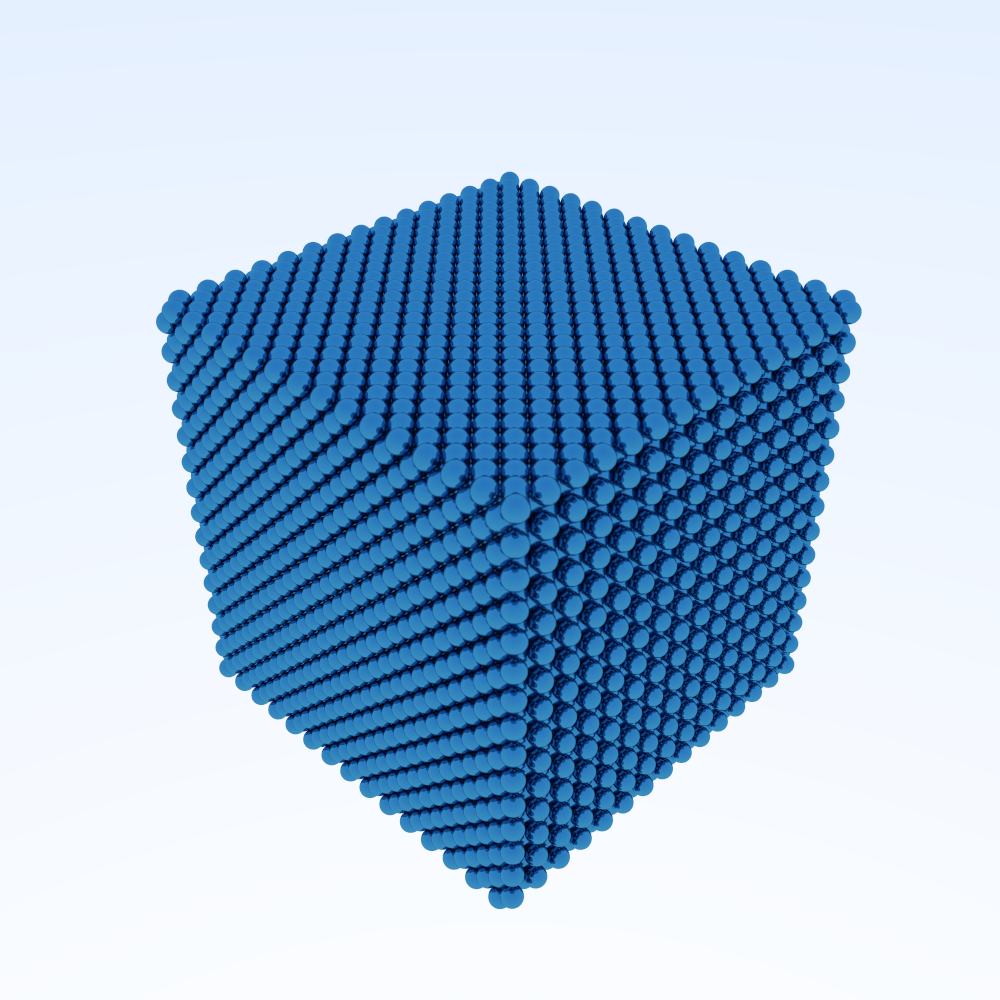}
    \includegraphics[width=0.48\columnwidth]{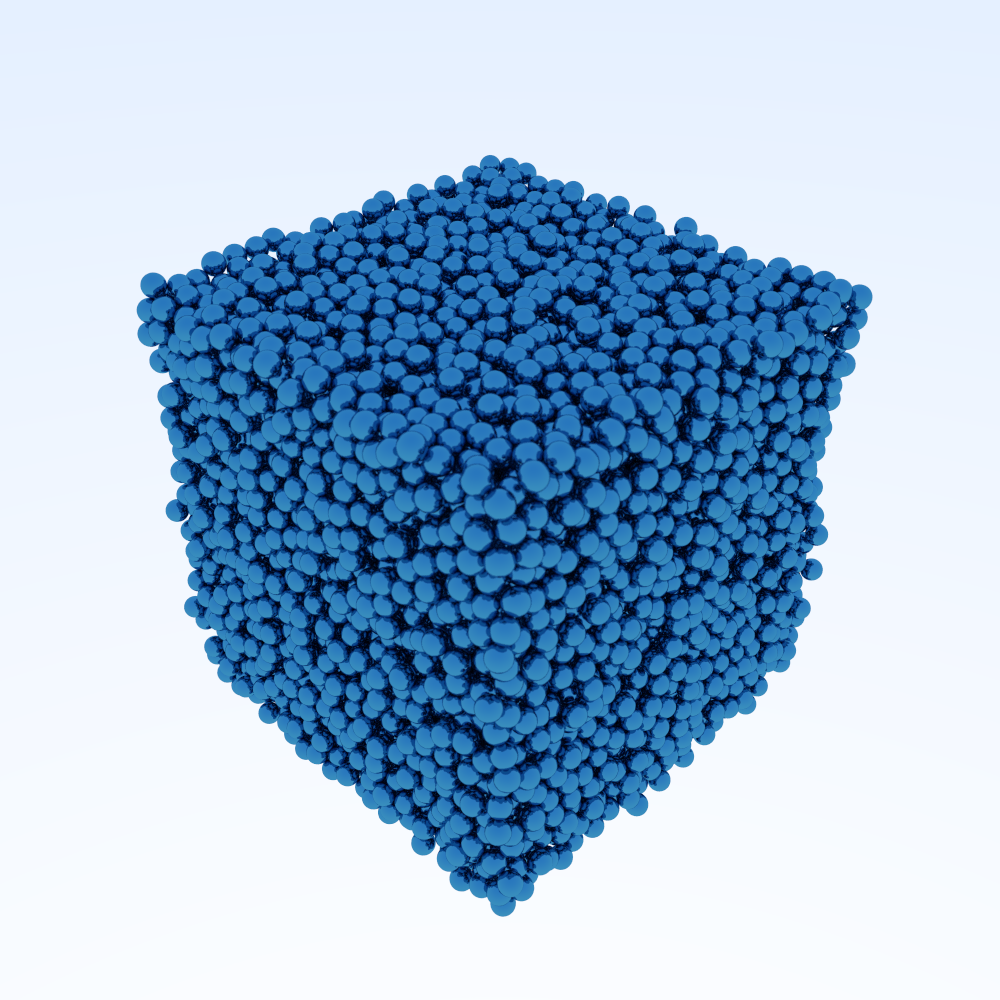}
    \caption{\textbf{Close packings of monodisperse hard spheres.}
    fcc configuration (left), and random packing with $\phi \approx 0.644$ (right) of monodisperse hard spheres.
    }
    \label{fig:MonoPackingsIllustration}
\end{figure}

When slowly compressing a hard-sphere fluid at constant temperature, a thermodynamically stable liquid branch can be defined from the ideal gas limit, $\phi=0$, until \textit{freezing}, $\phi_{\textrm{freeze}} \approx 0.494$. Further slow compression yields an entropy-driven first-order phase transition \citep{alder,wood,Hoover_Ree,Pusey_vanMegen} to a solid (crystalline) branch that extends from the \textit{melting} packing fraction, $\phi_{\textrm{melt}} \approx 0.545$, to the \textit{face-centered-cubic} (fcc) close-packing, $\phi_{\textrm{fcc}} = \frac{\pi}{6} \sqrt{2} \approx 0.7405$, shown in Fig.~\ref{fig:MonoPackingsIllustration}.
As already predicted by Kepler in his conjecture \cite{Torquato_Stillinger_Rev_Mod_2010} and formally proved by Hales \citep{Hales1,Hales2,Hales3}, the fcc crystal coincides with the densest ordered arrangement of hard spheres in $3d$. In this arrangement, the pressure diverges since the system cannot be further compressed.

It is well-known that by compressing hard spheres quickly, crystallization can be avoided \cite{Zaccarelli0,Zaccarelli1,Zaccarelli2}, so that the maximum close-packing density, $\phi_{\textrm{fcc}}$, is not reached and instead the particles ``jam'' in a disordered configuration at a lower volume fraction.
Just like the fcc crystal, these jammed configurations exhibit a diverging pressure, as further compression would lead to overlaps or deformation \citep{alfons}.
The determination of the so-called \textit{random close packing} (RCP) density, $\phi_{\textrm{RCP}}$, defined as the highest packing fraction for a ``disordered'' arrangement of hard spheres, remains an open problem \citep{van_Hecke_2009,Liu_Arxiv,Torquato_Stillinger_Rev_Mod_2010}.
In a classic experiment \citep{Bernal}, Bernal and Mason found that when equally sized spheres are poured and shaken in a container they occupy a volume fraction $\phi_{\textrm{RCP}} \approx 0.64$, a number that they conjectured to be ``mathematically determinable''.
An example of such a packing is shown in Fig.~\ref{fig:MonoPackingsIllustration}. Since then, measurements of $\phi_{\textrm{RCP}}$ have been reproduced in a myriad of experiments and numerical simulations, yet there is no consensus as to what the precise definition of RCP is~\cite{Kamien_Liu, Torquato_Stillinger_Rev_Mod_2010, Parisi_Zamponi_JCP, Wilken2021}.

In both experiments and in simulations, dense amorphous packings of hard spheres are produced by nonequilibrium dynamical processes, whose states are challenging to predict analytically~\citep{Krzakala_Kurchan,Torquato_dynamics}.
To overcome this difficulty, many authors have proposed that the RCP states correspond to the infinite-pressure limit of metastable glassy states~\citep{Parisi_Zamponi,Hermes_2010,mari,Berthier_Witten,speedy,Biazzo}, thus reducing a dynamical problem into a much simpler equilibrium one.
According to this view, when compressing a hard-sphere liquid beyond the freezing packing fraction, $\phi_{\textrm{freeze}}$, the pressure of the system first follows a metastable extension of the liquid branch and then becomes trapped in a glassy state, an amorphous solid state in which particles vibrate around random reference positions.
Upon further compression, the amplitudes of the vibrations eventually vanish and the pressure diverges as the system jams in a random packing.
Simulations showed that, depending on the compression rate, several glassy branches can arise from the metastable continuation of the liquid branch above $\phi_{\textrm{freeze}}$, and that different glasses can jam at different jamming densities~\citep{Hermes_2010, Berthier_Witten}.
Simulations and mean-field-level theory~\cite{Ozawa2017,Charbonneau2017,Parisi_Zamponi} indicate that these jamming densities live in a finite interval, between a lower bound obtained by compressing the least stable glassy branch, and an upper bound, usually called the glassy close-packing (GCP) density, defined as the densest possible jam with a glassy structure.

Alternative ways of thinking about random packings of hard spheres have been proposed by Torquato, Stillinger and co-workers \citep{Torquato_Truskett_Debenedetti_PRL,Rintoul_PRL,Rintoul_JCP}, Kamien and Liu \citep{Kamien_Liu}, and most recently Wilken et al. \citep{Wilken2021}.
Torquato, Stillinger and collaborators argued that the mechanical (compression) route to RCP is ill defined because one can always increase the volume fraction by locally ordering the particles~\citep{Torquato_Truskett_Debenedetti_PRL,Truskett,Kansal}.
Motivated by this observation, they introduced the alternative notion of a \textit{maximally random jammed} (MRJ) state corresponding to some minimum value of a structural order parameter, such as bond-orientational order \citep{Steinhardt1983}.
Adopting this criterion in numerical simulations, Rintoul and Torquato \citep{Rintoul_PRL,Rintoul_JCP} measured precise values of the pressure for the hard-sphere system on the metastable continuation of the liquid branch above the freezing point, $\phi_{\textrm{freeze}}$.
They found no evidence of thermodynamically stable amorphous (glassy) states, and observed a diverging pressure at $\phi_{\textrm{RCP}} \approx 0.644$ \citep{Rintoul_PRL,Rintoul_JCP}.

Kamien and Liu~\citep{Kamien_Liu}, conjectured a different definition for $\phi_{\textrm{RCP}}$ as the endpoint of the metastable extension of the equilibrium liquid branch.
More precisely, they linked the rate at which accessible states disappear to the pressure of the metastable liquid, and found that they both to diverge at $\phi_{\textrm{RCP}} \approx 0.64$, in accordance with previous numerical fits of the divergence of the pressure of the liquid branch~\cite{LeFevre1972,LeFevre1973}. In both approaches, $\phi_{\textrm{RCP}}$ is identified with the infinite pressure limit of a continuation of the equilibrium liquid branch, in agreement with ideas of Aste and Coniglio \citep{Aste_2004} and with recent work by Katzav \textit{et al.} \citep{Katzav}.

Finally, recent work by Wilken \textit{et al.} has proposed that RCP could be found as a dynamical critical point in an absorbing-state model, Biased Random Organization (BRO)~\cite{Wilken2021}.

\begin{figure}
    \centering
    \includegraphics[width=0.9\columnwidth]{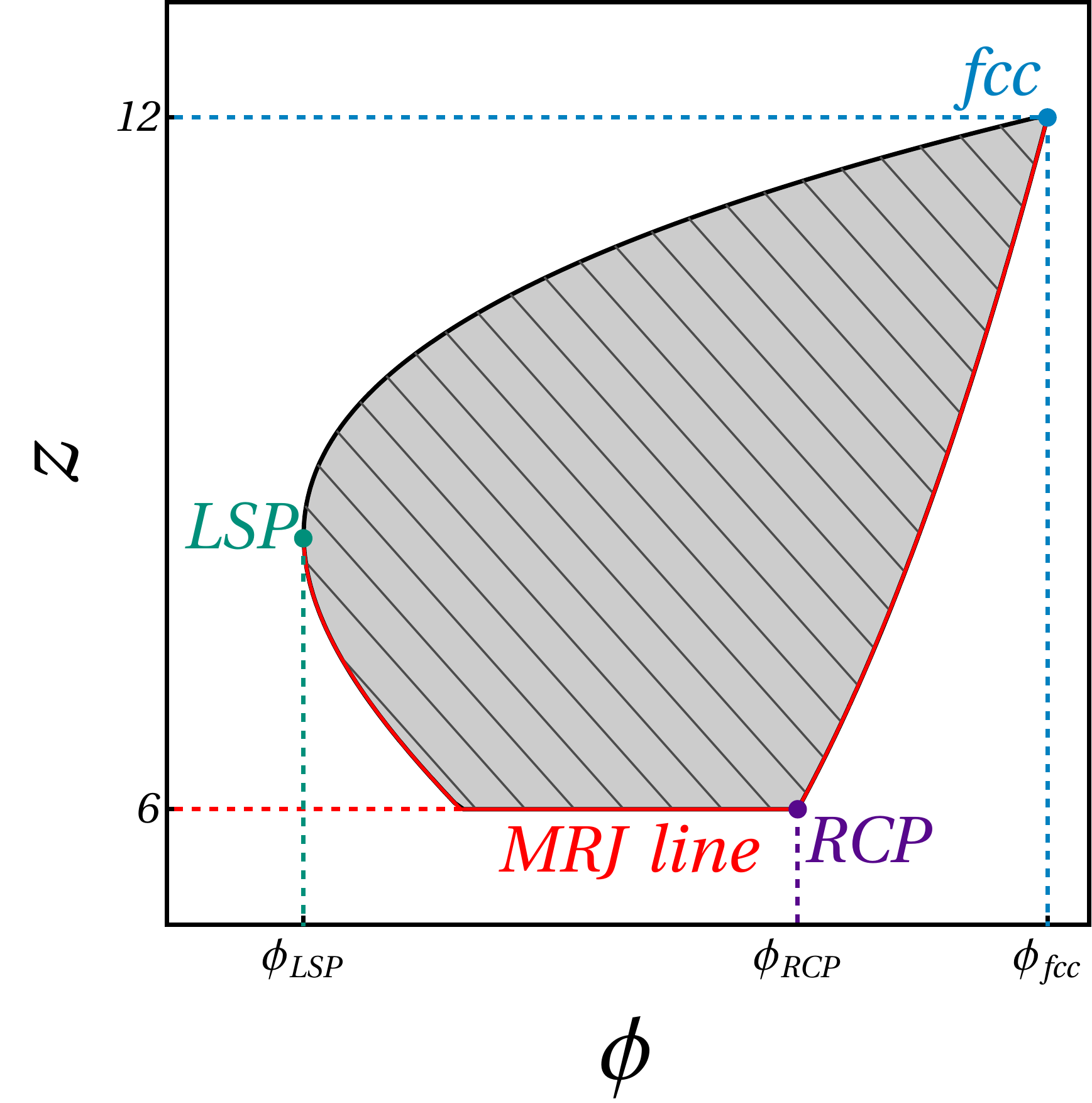}
    \caption{\textbf{Defining RCP in the $(\phi, z)$ plane.}
    Sketch of the convex hull of the ensemble of stable frictionless jammed packings (hashed gray region), in the packing fraction-kissing number plane.
    The least coordinated jam at every density form the MRJ-line for kissing numbers (solid red line).
    It starts at the loosest stable packing, here called LSP (teal dot), and ends at the densest possible packing fraction, fcc (blue dot).
    Between these extreme densities, there is a plateau of isostatic packings, which ends at a finite value, that we here use as a definition of RCP (purple dot).
    Past RCP, the kissing number picks up and reaches $12$ at fcc.}
    \label{fig:MRJLineSketch}
\end{figure}
We propose to define $\phi_{\textrm{RCP}}$ as a special point in the ensemble of jammed states, defined as follows, and sketched in Fig.~\ref{fig:MRJLineSketch}.
In the Torquato-Stillinger picture, for each density at which stable jammed states exist, one can rigorously define a conditional maximally random jammed state with respect to a given observable $\psi$.
This is an extension of the usual concept of MRJ, which defines a single density, to a whole MRJ-line (solid red line in Fig.~\ref{fig:MRJLineSketch}), extending from the density of the loosest stable packing (LSP), which is generally assumed to be part of a small family of defective crystalline states~\cite{Torquato2007,Torquato_Stillinger_Rev_Mod_2010}, all the way to the densest packing, fcc.
A particularly simple choice of observable (sometimes used in the MRJ picture~\cite{Jiao2011,Atkinson2014}) is the average number of contacts, or \textit{kissing number}, $z.$
In addition to being convenient, this choice is physically motivated by the fact that all the interpretations of RCP given above agree on the fact that RCP should be a point in the ensemble of jammed states where rigidity vanishes or, equivalently \cite{Zaccone_ScossaRomano}, where the packing is \textit{isostatic}, $z = 6$ in $3d$.
One can then seek a special density, that we shall henceforth call RCP, as the \textit{densest isostatic jammed packing}, i.e., the right-most point on the MRJ line in Fig.~\ref{fig:MRJLineSketch}.

Since the ``minimally coordinated" jammed packings for each density (i.e., on the MRJ-line) are in principle those closest in structure to liquid states, we adopt the viewpoint of Ref. \citep{Zaccone_PRL} and model the kissing number by well-known analytical approximations for the equation of state of a liquid, thereby invoking an analogy between the crowding of liquid and jammed states.
Like in all simple calculations, we make an assumption (here about crowding) that is wrong in detail, but we show that it captures critical aspects of the physics thus leading to nontrivial predictions that we validate by comparison with simulations and experiments.

\begin{figure}
    \centering
    \includegraphics[width=0.96\columnwidth]{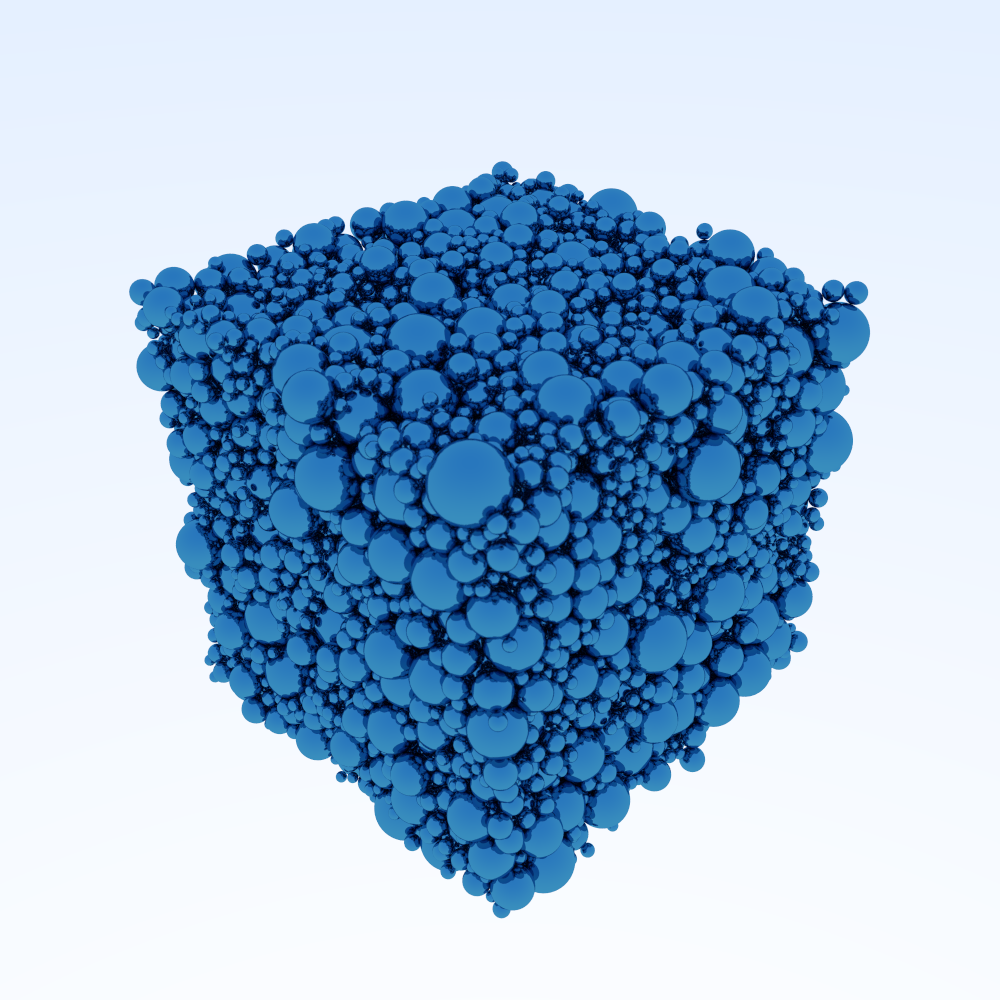}
    \caption{\textbf{Random packing of polydisperse hard spheres.}
    Random packing of polydisperse hard spheres with diameters drawn from the truncated power law considered in this paper, $(s_\sigma \approx 0.5, \phi\approx 0.719)$.}
    \label{fig:PolyPackingsIllustration}
\end{figure}

In the following, Sec.~\ref{sec:Theory}, we back the picture presented above in the case of monodisperse jammed packings, by showing that the number of contacts, $z$, empirically observed on the MRJ-line qualitatively agree with predictions from the ansatz of Ref.~\citep{Zaccone_PRL}.
Then, taking advantage of known extensions of liquid-state equations of state to polydisperse systems, we extend the framework of Ref.~\citep{Zaccone_PRL} to predict the value of $\phi_{\textrm{RCP}}$ as a function of polydispersity in hard-sphere fluids in $3d$.
While it is well known that polydisperse systems may pack to higher volume fractions than monodisperse systems (see example in Fig.~\ref{fig:PolyPackingsIllustration}), deriving good approximations for the values for $\phi_{\textrm{RCP}}$ as a function of the size polydispersity is not only of theoretical interest, but also of practical importance since these predictions can be used to guide experiments \cite{Chaikin}.

In Sec.~\ref{sec:discrete}, we show that our approach is reasonable for discrete distributions of particle sizes, using the example of a bidisperse mixture.  Since this system has been studied extensively, we compare our predictions to simulations of our own, as well as to data from a number of past computational~\citep{Biazzo,MENG_BINARY_MIXTURE, Farr_Groot_BINARY_MIXTURE,Kyrylyuk_BINARY_MIXTURE} and experimental~\cite{Yuan_BINARY_MIXTURE} works.
Then, in Sec.~\ref{sec:continuous}, we extend our approach to continuous polydispersities.
We assume the diameter of the spheres to follow three different size distributions, which have been widely employed to describe polydisperse colloidal suspensions in numerical simulations \citep{groot,Hermes_2010,Berthier_distribution1,Berthier_distributionPRX}.
We start by assuming the particle diameter to follow a log-normal distribution \citep{cramer_BOOK}, for which results from numerical simulations are available in the literature~\cite{groot}.
We then consider the particle diameter to follow a Gamma distribution, also known as Schulz distribution in this context~\cite{Schultz}, and a truncated power-law distribution recently introduced by Berthier and co-workers \citep{Berthier_distribution1, Berthier_distributionPRX}.
In all three cases, we show that $\phi_{\textrm{RCP}}$ increases monotonically with the relative standard deviation $s_{\sigma}$ of the distribution.
We compare the theoretical predictions both to data from the literature and to our own simulations. 
Finally, by a perturbative expansion we arrive at a closed form solution that captures a distribution independent regime for relative standard deviation $s_\sigma < 0.5$, and perform an analysis showing that the gradual loss of agreement for $s_\sigma>1$ can be associated with the growth of skewness in the distributions.

We end by drawing our conclusions in Sec~\ref{sec:Conclusions_and_outlook}.

\section{Theory \label{sec:Theory}}

\subsection{Monodisperse systems}

A property of random jammed states is that they are rigid, meaning that they exhibit a positive shear modulus, $G$.
For a disordered $d$-dimensional (with $d=2,3$) system of compressible spheres, the shear modulus can be shown to grow with coordination number, $z$, as $G \sim z - 2d$ \citep{Zaccone_ScossaRomano}.
Thus, for this class of systems, mechanical stability arises at a \textit{critical} coordination number $z_{c} \equiv 2d$, in agreement with Maxwell's isostaticity criterion.
The system is fluid for $z<z_{c},$ and jammed for $z \ge z_{c}$.

The value $z=6$ for hard spheres at RCP was independently reported in various contexts.
It was advanced as a result of analytical predictions stemming from the replica method~\cite{Parisi_Zamponi_JCP}.
Numerical simulations of fast compressions of finite-pressure glassy states confirmed this result over the whole range of replica-symmetry-breaking jammed states, that lie on the so-called ``J-line"~\cite{Ozawa2017,Charbonneau2017}.
Isostaticity was also empirically observed in simulations aiming to reach RCP while resorting to various dynamical processes unrelated to glassy physics~\cite{Wilken2021,Torquato2018}.
In this context, isostaticity has been observed in correspondence with the \textit{hyperuniformity} of the disordered sphere packings, meaning that long-range density fluctuations become anomalously suppressed or, equivalently, that the structure factor vanishes at small wavevectors as $S(|\mathbf{k}|\to 0) = |\mathbf{k}|^\alpha$, with $\alpha \approx 1/4$~\cite{Hexner2018,Wilken2021}.
Since hyperuniformity has been proposed as a prerequisite of RCP~\cite{Torquato2018}, the observation that hyperuniformity and $z = 6$ were observed at the same time lends credence to the validity of the isostaticity criterion.

Thus, using isostaticity, $z_{c} \equiv 6$, as a necessary (but not sufficient) criterion for RCP, we seek an ansatz for $z(\phi)$ along the MRJ line of the jammed domain. 
To this end, we introduce the radial distribution function (RDF), $g(r)$, representing the probability of finding (the center of) a particle in a shell of thickness $dr$ at a radial distance $r$ from (the center of) a test particle placed at the origin of the reference frame \citep{Hansen_McDonald_BOOK}.
By definition of the RDF, the average number of spheres lying in the range $r + d r$ is given by $d z = 4 \pi \rho g (r) r^{2} d r$.
By introducing the quantity $\sigma^{+} \equiv \sigma + \epsilon,$ where $\epsilon\to 0^{+}$ is an arbitrarily small number, the average number of particles in contact with (just touching) the test particle is given by
\begin{equation} \label{z_original}
z = 4 \pi \rho \int_{0}^{\sigma^{+}} g(r) r^{2} dr.
\end{equation}
The key point of the method introduced in Ref. \citep{Zaccone_PRL} is to treat $f(r) = 4 \pi \rho g (r) r^2/(N-1)$ as a \textit{partially continuous} probability distribution function (PDF).

In probability theory, besides fully continuous and fully discrete PDFs, one can define partially continuous distributions, also known as \textit{mixed distributions} \citep{shynk_BOOK}.
As an example of a fully discrete distribution, the PDF $f_{d} (x)$ of a distribution consisting of a set of possible outcomes $x_{i} = \left\{ x_{1}, \cdots, x_{n} \right\}$ with corresponding probabilities $p_{i} = \left\{ p_{1}, \cdots, p_{n} \right\},$ can be written as $f_{d} (x) = \sum_{i=1}^{n} p_{i} \delta (x - x_{i}).$
A partially continuous (PC) distribution can be written as \citep{shynk_BOOK}: $f_{\textrm{PC}} (x) = c(x) + \sum_{i=1}^{n} p_{i} \delta (x - x_{i})$ where $c(x)$ is the continuous part and the second term is the discrete part. Upon normalizing to $1$ over the relevant domain, $\int_{0}^{\infty} f_{\textrm{PC}} (x) dx = 1,$ $f_{\textrm{PC}} (x)$ becomes a valid PDF \citep{pishro_BOOK}.

\begin{figure}
  \centering
  \includegraphics[width=6.5cm]{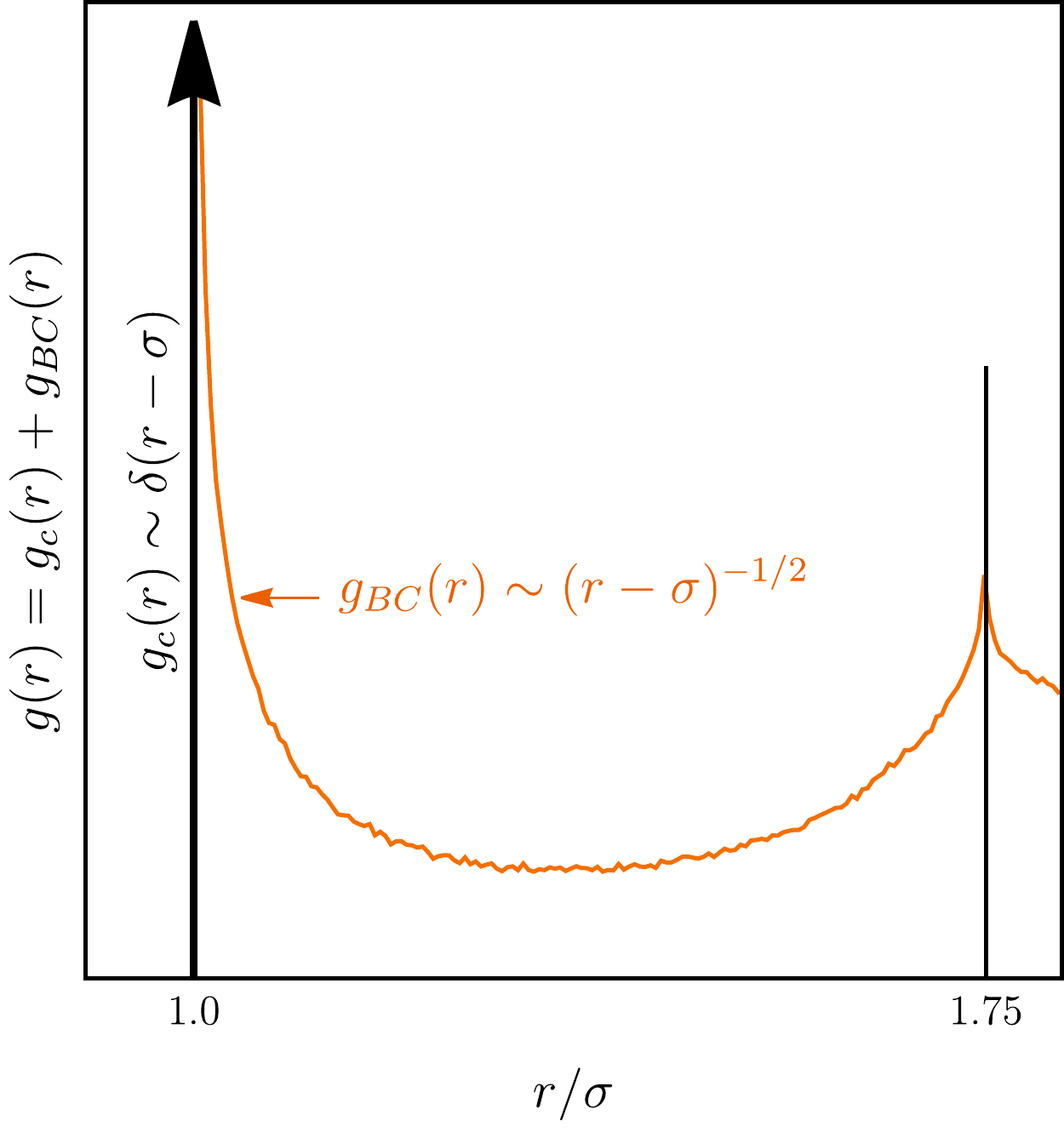}
  \caption{
  \textbf{Partially continuous RDF.}
  Radial distribution function (RDF) $g(r)$ of a system of hard spheres with diameter $\sigma$ in $d=3$ dimensions, at the random close packing density $\phi_{\textrm{RCP}}.$ 
%   The black curve representing numerical data adapted from Ref. \citep{Donev},
    The orange curve, representing data obtained in the simulations introduced hereafter, is the \textit{continuous}  part $g_{\textrm{BC}} (r)$ of the RDF.
    We here indicate the exponent from Ref. \citep{Donev} (note that a slightly different exponent $0.4$ was discovered in Ref. \citep{Lerner} and appears in agreement with replica theory \citep{Charbonneau_JSTATPHYS}, distinguishing between these values is beyond the scope of this paper).
    The thick vertical arrow represents the Dirac delta in the \textit{discrete} part $g_{c} (r)$ of the RDF (see Eq. \eqref{g_c}).}
 \label{Fig:partialg}
\end{figure}
In short, we write the RDF as
\begin{equation} \label{g_partially_continuous}
g(r) = g_{c} (r) + g_{\textrm{BC}} (r),
\end{equation}
where $g_{\textrm{BC}} (r)$  is the \textit{continuous} part describing the probability of finding particles in the region of space beyond contact (BC) $r > \sigma^{+},$ while $g_{c} (r)$ is the \textit{discrete} part describing the probability of having nearest neighbors in direct contact with the test particle.
We then write $g_{c} (r)$ as
\begin{equation} \label{g_c}
g_{c} (r) = g_{0} g (\sigma; \phi) \delta (r- \sigma),
\end{equation}
where $g_{0}$ is a normalization length, while $g (\sigma; \phi)$ is the so-called (dimensionless) \textit{contact value} of the RDF at packing fraction $\phi$, and represents the probability of finding particles at exactly $r= \sigma$. 
The total $g(r)$ given by Eq. \eqref{g_partially_continuous} obeys the usual condition $\int_{0}^{\infty} 4 \pi \rho g(r) r^{2} dr =N - 1$, imposed by normalization.
This separation of the $g(r)$ of jammed states of hard spheres into a continuous and a discrete part, illustrated in Fig.~\ref{Fig:partialg}, is consistent with the previous works~\cite{Dimon,LATTUADA,Torquato_2018}.

Upon insertion of Eq. \eqref{g_c} into Eq. \eqref{g_partially_continuous} and the resulting expression into Eq. \eqref{z_original}, the coordination number $z$ arising from the particles in permanent contact with the test particle is given by
\begin{equation} \label{zc_condition}
z = 24 \phi \frac{g_{0}}{\sigma} g (\sigma; \phi).
\end{equation}
If $g_{0}/\sigma$ and $g (\sigma; \phi)$ were known on the branch of maximally random jammed states, the RCP density $\phi_{\textrm{RCP}}$ could be found by solving Eq. \eqref{zc_condition} while imposing the critical condition for the onset of mechanical stability $z=z_{c} \equiv 6.$ 
However, jammed states are notoriously hard to model due to their non-equilibrium nature.
In order to use Eq.~ \eqref{zc_condition} to predict RCP, in the absence of a better theory, we introduce an analogy with equilibrium, that has also the benefit of yielding analytically tractable equations.

In equilibrium hard spheres, due to the virial theorem, the value $g_{\textrm{eq}} (\sigma; \phi)$ of the RDF at contact provides the pressure $p$ of the uniform fluid as a function of its packing fraction $\phi  \equiv \frac{4}{3} \pi (\sigma/2 )^{3} \rho, $ through the relation
 \citep{Allen_Tildesley,Torquato_BOOK,Hansen_McDonald_BOOK}
\begin{equation} \label{Z_mono}
Z(\phi) = 1 + 4 \phi g_{\textrm{eq}}(\sigma;\phi),
\end{equation}
where $Z \equiv p / \rho k_{\textrm{B}} T$ is the so-called compressibility factor, $\rho$ is the number density, $T$ and  $k_{\textrm{B}}$ are the temperature and the Boltzmann constant, respectively.
This expression (that is exact for equilibrium liquids) can of course not be used directly for jammed states. In particular, $g_{\textrm{eq}}$ is a regular function of $r$ for all $r > \sigma$, so that $g_{\textrm{eq}}(\sigma; \phi)$ is fundamentally different from $g(\sigma; \phi)$, the amplitude of the singular part of the jammed RDF at contact.
This difference is consistent with the fact that the pressure has to diverge in collectively jammed states \citep{Torquato_Truskett_Debenedetti_PRL,Rintoul_PRL,Rintoul_JCP,Kamien_Liu,Aste_2004}.

By \textit{analogy} with equilibrium states, to qualitatively describe local crowding in hyperstatic, maximally random jammed states, we propose to write
\begin{equation} \label{g_analogy}
g(\sigma; \phi) \propto g_{\textrm{eq}}(\sigma; \phi) = \frac{Z(\phi) - 1}{4\phi},
\end{equation}
with $Z$ an approximate analytical equation of state of equilibrium hard spheres.
We list the expressions for $Z$ used in this paper in App.~\ref{app:EoS}.
Injecting Eq.~\eqref{g_analogy} into Eq.~\eqref{zc_condition} yields
\begin{equation} \label{zc_condition_ansatz}
z = 6 \phi C_0 \left(Z(\phi) - 1\right),
\end{equation}
where $C_0 \equiv g_{0}/ \sigma$ is a constant number to be determined.
Intuitively, this ansatz assumes that the most random branch of jammed states undergoes crowding in a way that would be qualitatively similar to an equilibrium liquid.

The last ingredient needed to solve Eq. \eqref{zc_condition_ansatz} and thus to find an expression for $\phi_{\textrm{RCP}},$ is the value of the constant factor $C_{0}$.
To fix its value, we insert in Eq. \eqref{zc_condition_ansatz} a known $(\phi_{\mathrm{ref}},z_{\mathrm{ref}})$ combination, typically from a perfect crystalline packing, as well as a choice of equation of state.
This procedure can be seen as an effective ``boundary condition'' in our problem.
In Ref. \citep{Zaccone_PRL}, the author chose fcc ordering, \textit{i.e.} a coordination number $z_{\textrm{fcc}}=12$ and a packing fraction $\phi^{\textrm{CP}}_{\textrm{fcc}} =  \pi / 3 \sqrt{2} \approx 0.74$ \cite{Torquato_Stillinger_Rev_Mod_2010}.
This choice is justified by the picture that maximally random jammed states have to connect RCP to the fcc point, see Fig.~\ref{fig:MRJLineSketch}.
Another suggestion~\citep{likos} has been to use perfect bcc ordering, identified by the coordination number $z_{\textrm{bcc}}=8$ and packing fraction $\phi^{\textrm{CP}}_{\textrm{bcc}} =  \pi \sqrt{3}/8 \approx 0.68$.

\begin{figure}
    \centering
    \includegraphics[width=0.9\columnwidth]{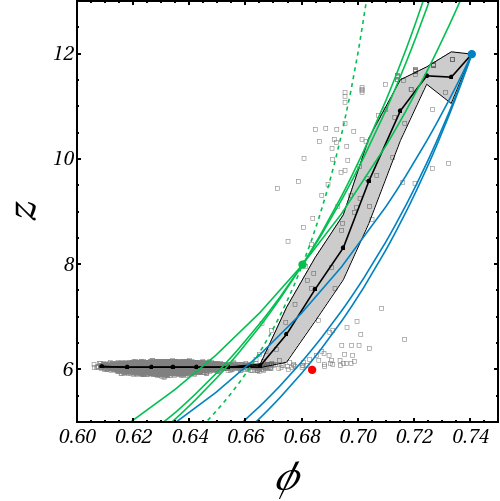}
    \caption{
    \textbf{Kissing number of maximally random jams.}
    Scatter plot of the kissing number against the final packing fraction for $10^5$ compressions of $N = 108$ particles, using a Torquato-Jiao algorithm.
    Each open gray square represents one compression.
    Black disks are binned averages, and the gray area represent the confidence interval around it.
    Colored disks represent three known special points: fcc (blue), bcc (green), and the Parisi-Zamponi~\cite{Parisi_Zamponi} estimate of GCP in $d=3$ (red).
    Colored lines represent our ansatz, when using fcc (blue) or bcc (green) boundary conditions to set $C_0$.
    Solid lines were obtained, from left to right, using the PYv, CS, and PYc equations of state.
    The dashed lined was obtained using the Young-Alder equation of state.
    }
    \label{fig:z_versus_phi}
\end{figure}
In order to check how reasonable our ansatz is, we generate $10^5$ jammed packings of $N = 108$ particles using the Torquato-Jiao algorithm~\cite{Jiao2011}, which was designed to generate strictly jammed packings that are as random as possible (See App.~\ref{app:numerics} for details on the algorithm).
The reason for using a small number of particles and a large number of compressions is that the distribution of final jammed densities of such compression algorithms is typically heavily peaked around $\phi = 0.64$, so that measuring configurations in the regime leading up to fcc requires a lot of compressions.
At the end of each compression, we measure the average kissing number in the system, as well as the final packing fraction, and we report these values in Fig.~\ref{fig:z_versus_phi}.
The lowest jammed densities are obtained at roughly $\phi \approx 0.6$, as reported in previous works~\cite{Jiao2011}, while the densest packings are found at the fcc density.
As described in similar simulations of hard disks~\cite{Atkinson2014}, a roughly flat region indicates that only isostatic packings are found at the lowest observed jammed densities.
Between these two regimes, the kissing number picks up, joining the $z = 6$ region and the $z=12$ point.

To give a better idea of the statistics of points within the scatter plot in Fig.~\ref{fig:z_versus_phi}, we show binned averages as a black line, and binned standard deviations as a grey area.
While only rare fluctuations around isostaticity are observed up to $\phi \approx 0.67$, the average kissing number picks up after that value, with a rather large spread until fcc, that could be attributed to finite-size effects.
On this plot, we also represent three special points as colored disks: fcc, in blue at $(\pi/\sqrt{18},12)$, bcc, in green at $(\pi \sqrt{3}/8, 8)$, and the $3d$ value of the Glass Close-Packing (GCP) predicted by mean-field theory~\cite{Parisi_Zamponi}, at $(0.6836, 6)$.
Note that the GCP point roughly matches with the point where the lower bound of the scatter plot picks up from $z = 6$.
Moreover, the bcc point seems to lie on the upper limiting curve around the observed points.
Finally, we plot predictions from our ansatz as solid lines, in blue when fcc is used to determine $C_0$, and in green when bcc is used instead.
The solid lines are obtained using (from left to right on the plot) the PYv, CS, and PYc equations of state, while the dashed line was obtained using the YA expression (see App.~\ref{app:EoS} for their expressions).

These different predictions are spread in a rather broad region, but they follow the right qualitative trend compared to data -- which was not guaranteed, since the analytical equations of state used to draw them are not supposed to describe this regime.
For each boundary condition and equation of state, a value of $C_0$ as well as a closed-form expression for $\phi_{\textrm{RCP}}$ can be obtained. The obtained values are summarised in Table \ref{table1}. 

\begin{table}[b]
\centering % used for centering table
\begin{tabular}{c c c c c} % centered columns (4 columns)
\hline\hline %inserts double horizontal lines
fcc & PYv & \ CS & \ PYc & \ YA \\ [0.5ex] % inserts table
%heading
\hline % inserts single horizontal line
$ 10^{2} \cdot C_0$ & 3.31894 & 1.87416 & 1.53909 & N/A \\ % inserting body of the table
$\phi_{\textrm{RCP}}$ & 0.658963 & 0.677376 & 0.68086 & N/A \\ [1ex] % [1ex] adds vertical space
\hline %inserts single line
bcc & PYv & \ CS & \ PYc & \ YA \\ [0.5ex] % inserts table
%heading
\hline % inserts single horizontal line
$ 10^{2} \cdot C_0$ & 3.74068 & 2.42946 & 2.06716 & 3.73673 \\ % inserting body of the table
$\phi_{\textrm{RCP}}$ & 0.643320 & 0.650594 & 0.652187 & 0.660868 \\ [1ex] % [1ex] adds vertical space
\end{tabular}
\caption{Normalization factor $C_0$ (see Eq. \eqref{zc_condition_ansatz}) and random close packing density $\phi_{\textrm{RCP}}$ of a monodisperse fluid of hard spheres with diameter $\sigma,$ in $d=3$ dimensions, obtained for different approximations (Percus-Yevick with either the virial (PYv) or compressibility (PYc) equation of state, Carnahan-Starling (CS), and Young-Alder (YA)) for the contact value $g (\sigma)$ of the radial distribution function, and different configurations (fcc or bcc) as boundary conditions. Note that the YA equation diverges at fcc, so that it is not usable with fcc as a reference.} 
\label{table1} % is used to refer this table in the text
\end{table}
Note that it is not clear at this stage whether any of these approximations is objectively better than the others, since there is no ground truth for the value of RCP, nor for the branch of interest of $z(\phi)$, which in the numerical measurements of Fig.~\ref{fig:z_versus_phi} is probably marred by finite-size effects.
More specifically, all tested equations of state yield values in a reasonable interval compared to the literature~\footnote{Commonly cited values are $\phi_{\textrm{RCP}} \approx 0.642-0.649$ by finite-rate compression compression~\cite{Jodrey1985,Jullien1997}, $0.68$ by a Monte Carlo method~\cite{Tobochnik1988}, $0.64$ by differential-equation densification~\cite{Zinchenko1994}, $0.60$ by ``drop and roll''~\cite{Visscher1972}, $0.64-0.65$ by the LS algorithm and its variants~\cite{Torquato_Stillinger_Rev_Mod_2010,Baranau2014}, and $0.64$ by Biased Random Organization~\cite{Wilken2021}.}, suggesting that models of $z(\phi)$ that travel close to fcc and bcc would also yield reasonable values.
For instance, as far as the value of the monodisperse RCP density alone is concerned, one could also use a completely unphysical fit for $z(\phi)$.
An extreme example of this would be, say, a linear approximation going through both fcc and bcc: this completely unjustified approximation would lead to yet another reasonable value in closed form, $\phi_{\textrm{RCP}} = \pi (9 \sqrt{3} - 4 \sqrt{2})/48 \approx 0.65$.

However, we shall show in the next section that there is a major advantage in using an actual equilibrium equation of state as a model for crowding.
Namely, since equations of states of monodisperse hard spheres have been extended to polydisperse hard spheres, there is a natural extension of this computation to polydisperse systems, which we shall show correctly captures the evolution of $\phi_{\textrm{RCP}}$ with increasing polydispersity.
 
\subsection{Polydisperse systems}

In order to extend this theoretical framework to polydisperse systems, we consider an $m$-component mixture of additive hard spheres in $d=3$ dimensions.
% The continuously polydisperse case is then easily obtained by taking $m \to \infty$.
We call  $\sigma_{ij} = \frac{1}{2} (\sigma_{i} + \sigma_{j})$ the contact distance between a sphere of species $i$ and a sphere of species $j,$ where $\sigma_{ii} \equiv \sigma_{i}$ is the diameter of a sphere of species $i.$
We indicate the number fraction of species $i$ with $x_{i}= \rho_{i}/\rho,$ where $\rho$ is the number density of the mixture while $\rho_{i}$ is the number density of spheres of species $i.$ 
Finally, we define $\left\langle \sigma^{n} \right\rangle \equiv \sum_{i=1}^{m} x_{i} \sigma_{i}^{n}$ such that the packing fraction of the system is given by $\phi = \pi \rho \left\langle \sigma^{3} \right\rangle /6.$ 

To predict the RCP density, $\phi_{\textrm{RCP}}$, of a mixture as we did above, Eq. \eqref{zc_condition} needs to be suitably modified.
The mean number of contacts, $z_{ij}$, between particles of species $i$ and those of species $j$ is linked to the partial RDF, $g_{ij}$, restricted to $ij$ pairs, through
\begin{align}
z_{ij} &= 4 \pi \rho \int\limits_{0}^{\sigma_{ij}^+} dr r^2 g_{ij}(r).
\end{align}
Like in the monodisperse case, the only part of $g_{ij}$ that participates in the kissing number is the contact value $g_{ij,c}$, so that
\begin{align}
    z_{ij} &= 24 \phi \frac{\sigma_{ij}^2}{\langle \sigma^3 \rangle} g_{ij,c}(\sigma_{ij}; \phi).
\end{align}
We then write the value of the \textit{species-averaged} kissing number, $\langle z \rangle$, as
\begin{align}
    \langle z \rangle  &= 24 \phi \sum\limits_{i,j}^{m} x_i x_j \frac{\sigma_{ij}^2}{\langle \sigma^3 \rangle} g_{ij,c}(\sigma_{ij}; \phi).
\end{align}
Finally, one needs to assume an expression for $g_{ij, c}$.
The latter should be $i\leftrightarrow j$ symmetric, and converge to its monodisperse expression, $g_0 g(\sigma; \phi)$, in the limit of a single species, which is attained either by enforcing that $m =1$, or by imposing that all diameters are equal, $\sigma_i = \langle \sigma \rangle$.
A simple functional form that verifies all of the above is 
\begin{align}
g_{ij,c}(\sigma_{ij}; \phi) \equiv \frac{\sigma_{ij}}{\langle \sigma \rangle} g_0(\langle \sigma \rangle) g_{ij}(\sigma_{ij}; \phi),
\end{align}
which yields the expression
\begin{align}
\langle z \rangle  &= 24 \phi \frac{g_0}{\langle\sigma\rangle} \sum\limits_{i,j}^{m} x_i x_j \frac{\sigma_{ij}^3}{\langle \sigma^3 \rangle} g_{ij}(\sigma_{ij}; \phi).
\end{align}
This last equation is consistent with known expressions of the compressibility factor $Z^{(m)}$ (and therefore the species-averaged pair correlation function at contact appearing in the virial theorem, $g^{(m)}_{\textrm{eq}}$) of equilibrium polydisperse hard spheres~\citep{lebowitz,MCSL,santos,mulero}
\begin{equation} \label{Z_mixture}
 g^{(m)}_{\textrm{eq}}(\sigma;\phi) \equiv \frac{Z^{(m)} (\phi) - 1}{4 \phi} =  \sum_{i=1}^{m} \sum_{j=1}^{m} x_{i} x_{j} \frac{\sigma_{ij}^{3}}{\left\langle \sigma^{3} \right\rangle} g_{ij} (\sigma_{ij}; \phi).
\end{equation}
All in all, the analogy between least-coordinated jammed packings and equilibrium fluids invoked in the monodisperse case naturally generalizes to the polydisperse case as
\begin{align} \label{condition_DICIOTTO_POLY}
\langle z \rangle  &= 6 C_0 (Z^{(m)} (\phi) - 1).
\end{align}
Furthermore, the mechanical stability criterion still requires isostaticity at the level of the average number of contacts, $\langle z\rangle = z_c \equiv 6$, so that the only change between monodisperse and polydisperse packings in our approach is the equilibrium equation of state used in the analogy.

This result can be further generalized to the case of a continuously polydisperse system of hard spheres whose diameters follow a continuous distribution $f(\sigma),$ by considering the limit $m \to \infty.$ 
In this case, Eq. \eqref{Z_mixture} becomes \citep{lado1996}
\begin{equation}
\begin{split}
& g^{(m \to \infty)}_{\textrm{eq}}(\sigma; \phi) \equiv \frac{Z^{(m \to \infty)} (\phi) - 1}{4 \phi} =\\
&\frac{1}{ 8 \left\langle \sigma^{3} \right\rangle } \int_{0}^{\infty} d \sigma \int_{0}^{\infty} d \sigma' f (\sigma) f (\sigma')  ( \sigma + \sigma')^{3} g (\sigma, \sigma'; \phi), \label{Z_poly}
\end{split}
\end{equation}
where now $\left\langle \sigma^{n} \right\rangle = \int_{0}^{\infty} d \sigma  f(\sigma) \sigma^{n}.$ 
Note that Eq. \eqref{Z_mixture} for the $m$-component mixture can be recovered by taking $f (\sigma) = \sum_{i=1}^{m} x_{i} \delta (\sigma_{i} - \sigma).$

The protocol used in this paper to compute the RCP density, $\phi_{\textrm{RCP}}$, of a polydisperse hard-sphere system then goes as follows. We use an approximate expression for the EOS, $Z^{(m \to \infty)} (\phi)$, of the system under study, which yields an estimate of $g^{(m \to \infty)}(\phi)$ through Eq. \eqref{Z_poly}.
By analogy with the monodisperse case, we then find $\phi_{\textrm{RCP}}$ by substituting $Z^{(m \to \infty)} (\phi)$ into Eq. \eqref{condition_DICIOTTO_POLY} and imposing the critical condition for jamming, $\left\langle z \right\rangle = z_{c} \equiv 6.$
In other words, we solve
\begin{equation}
1 = C_0 \left(Z^{(m \to \infty)} (\phi_{\textrm{RCP}}) - 1\right),
\end{equation}
where, since Eq. \eqref{Z_poly} correctly reduces to Eq. \eqref{Z_mono} in the limit of a one-component system, we use the values in the upper rows of Table \ref{table1} for the normalization factor $C_{0}$.
The equations of states used in the polydisperse case are the Boubl\'{i}k-Mansoori-Carnahan-Starling-
Leland (BMCSL), extended Carnahan-Starling (eCS), and extended Percus-Yevick (ePY) equations. $Z_{\textrm{BMCSL}} (\phi),$ $Z_{\textrm{eCS}} (\phi)$ and $Z_{\textrm{ePY}} (\phi)$, are defined in App.~\ref{app:EoS}.

\subsection{Strategy recap \label{sec:recap}}

The strategy we propose to predict $\phi_{\textrm{RCP}}$ in a polydisperse hard-sphere system can be summarised as follows. First, given a size distribution $f(\sigma)$, we derive an approximate analytical EOS $Z^{(m \to \infty)} (\phi)$ from either Eq. \eqref{ZBMCSL} or \eqref{Zsantos}.
From this EOS, we deduce an estimate of the averaged (over the size distribution) contact value of the radial distribution function for the polydisperse system through Eq. \eqref{Z_poly},
\begin{align*}
    g^{(m \to \infty)}_{\textrm{eq}}(\sigma,\phi) \equiv \frac{Z^{(m \to \infty)} (\phi) - 1}{4 \phi}.
\end{align*}
Furthermore, we compute the value of $C_0$ using the monodisperse limit of the EOS, $Z(\phi)$, and some known combination of $\phi_{\mathrm{ref}}$ and $z_{\mathrm{ref}}$ for the monodisperse fluid, by solving
\begin{align*}
    C_0  =  \frac{z_{\mathrm{ref}}}{6 (Z(\phi_{\mathrm{ref}}) - 1) }.
\end{align*}
Finally, we insert $Z^{(m \to \infty)}$ and $C_0$ into Eq.~\eqref{condition_DICIOTTO_POLY}, and impose $\left\langle z \right\rangle = z_c \equiv 6$.
In the end, an estimate of $\phi_{\textrm{RCP}}$ is obtained by solving
\begin{equation}
1 = C_0 \left(Z^{(m \to \infty)} (\phi_{\textrm{RCP}}) - 1 \right).
\label{eq:SimplerCondition}
\end{equation}
Note that, since the compressibility factor of the liquid branch is typically a strictly growing function of the packing fraction, $Z^{(m \to \infty)}$ can be inverted and this equation admits a single solution.

\section{Results}

In this section, we present predictions for $\phi_{\textrm{RCP}}$ obtained using the framework of Sec.~\ref{sec:recap}, first in the case of discrete polydispersity, Sec.~\ref{sec:discrete}, then in the case of a continuous distribution of particle diameters, Sec.~\ref{sec:continuous}.
Our predicted values for each size distribution are compared to numerical data, some adapted from previous numerical work, and some obtained ourselves using the same method as in Ref.~\cite{Baranau2014}, namely a modified Lubachevsky-Stillinger~\cite{Lubachevsky1990,Lubachevsky1991} compression algorithm that enables to reach large packing fractions in random packings (see Appendix~\ref{app:numerics} for details of the simulations). Where available we also compare to experimental data~\citep{Yuan_BINARY_MIXTURE}.

\subsection{Discrete polydispersity (bidispersity)\label{sec:discrete}}

We start by assuming the particle diameter $\sigma$ to follow a discrete probability distribution.
More specifically, in order to compare our results with those present in the literature \citep{Biazzo,MENG_BINARY_MIXTURE, Farr_Groot_BINARY_MIXTURE,Kyrylyuk_BINARY_MIXTURE,Yuan_BINARY_MIXTURE}, we consider the particle diameter to follow a bidisperse distribution.
The system then contains $N_1$ spheres with diameter $\sigma_1$, and $N_2 = N - N_1$ spheres with diameters $\sigma_2$.
Introducing the number fraction of each species, $x_{1,2} \equiv N_{1,2}/N$, the corresponding size distribution can be written as
\begin{equation} \label{bidisperse_probability_BINARY_MIXTURE}
f (\sigma) = x_{1} \delta (\sigma_{1} - \sigma) +  x_{2} \delta (\sigma_{2} - \sigma).
\end{equation}
The $n$-th moment $\left\langle \sigma \right\rangle \equiv \int_{-\infty}^{\infty} d \sigma f(\sigma) \sigma^{n}$ of the probability distribution \eqref{bidisperse_probability_BINARY_MIXTURE} is given by 
\begin{equation} \label{moments_BINARY_MIXTURE}
\left\langle \sigma^{n} \right\rangle = x_{1} \sigma_{1}^{n} + x_{2} \sigma_{2}^{n}.
\end{equation}
Insertion of Eq. \eqref{moments_BINARY_MIXTURE} in the $Z_{\textrm{BMCSL}} (\phi),$ $Z_{\textrm{eCS}} (\phi)$ and $Z_{\textrm{ePY}} (\phi)$ introduced in the previous section, allows us to find three distinct approximate analytical expressions for the EOS $Z^{(m=2)} (\phi)$ of this bidisperse system.
The effect of the discrete polydispersity on the system can be fully described by the diameter ratio, $\sigma_{1} / \sigma_{2}$, and either one of the number fractions $x_{1,2}$.
It is however common in the literature to use the volume fractions of the species, $\eta_{2} \equiv x_{2}/ \big( x_{2} + x_{1} ( \sigma_{1}/ \sigma_{2})^{3} \big) $ and $\eta_1 = 1 - \eta_2$, instead of the number fraction.
For the rest of this section, we adopt the same convention: we denote the species with larger diameter with index $1$, so that $\sigma_{1} / \sigma_{2} > 1$, and plot the RCP density as a function of the volume fraction $\eta_2$ of the species with smaller diameter.
\begin{figure}
\centering
\includegraphics[width = 1.1 \linewidth]{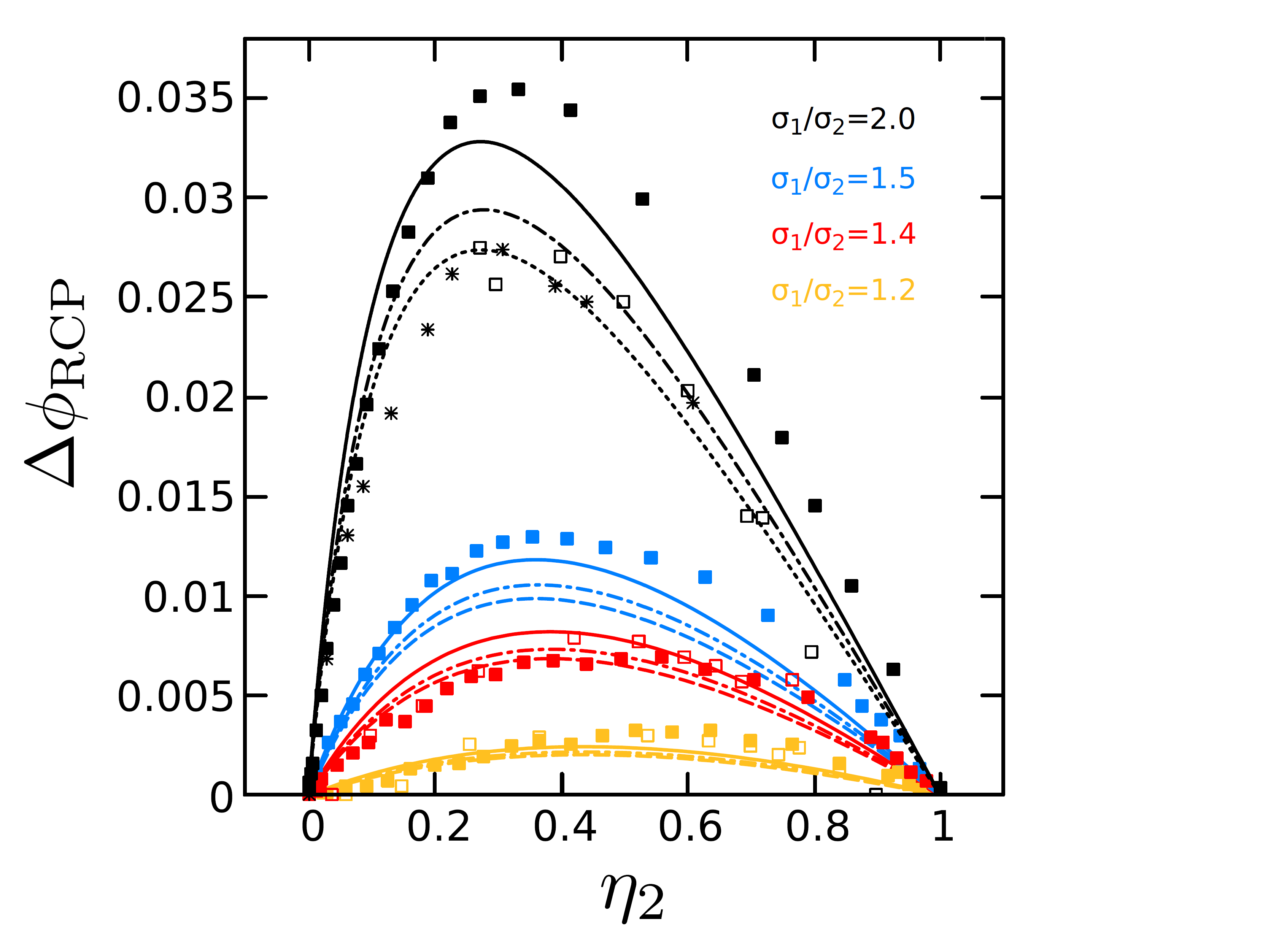}
  \caption{\textbf{Polydispersity-induced shift of RCP of a binary mixture of hard spheres.}
  Shifted random close packing density $\Delta \phi_{\textrm{RCP}} \equiv \phi_{\textrm{RCP}} - \phi_{\textrm{RCP}}^{\textrm{mono}}$  against the volume fraction of spheres with diameter $\sigma_{2},$ in a binary mixture with fixed diameter ratio $\sigma_{1} / \sigma_{2} =1.2$ (yellow), $\sigma_{1} / \sigma_{2} =1.4$ (red), $\sigma_{1} / \sigma_{2} =1.5$ (blue), and $\sigma_{1} / \sigma_{2} =2$ (black), respectively.
In all cases full, dashed and dot-dashed black lines represent results obtained when using the $Z_{\textrm{ePY}} (\phi),$ $Z_{\textrm{eCS}} (\phi)$ and $Z_{\textrm{BMCSL}} (\phi)$ approximations for the equation of state of the system, respectively, and the bcc configuration as a boundary condition to determine $C_{0}.$ Open points represent simulations from Ref.~\citep{Biazzo} while filled points are simulations of our own. Black star-shaped points are results recently obtained in Ref. \citep{Houfei_KOB} for a binary granular system.}
		\label{FIGURE_1_BINARY_MIXTURE}
\end{figure}

\begin{figure}
\centering
\includegraphics[width = 1.0 \linewidth]{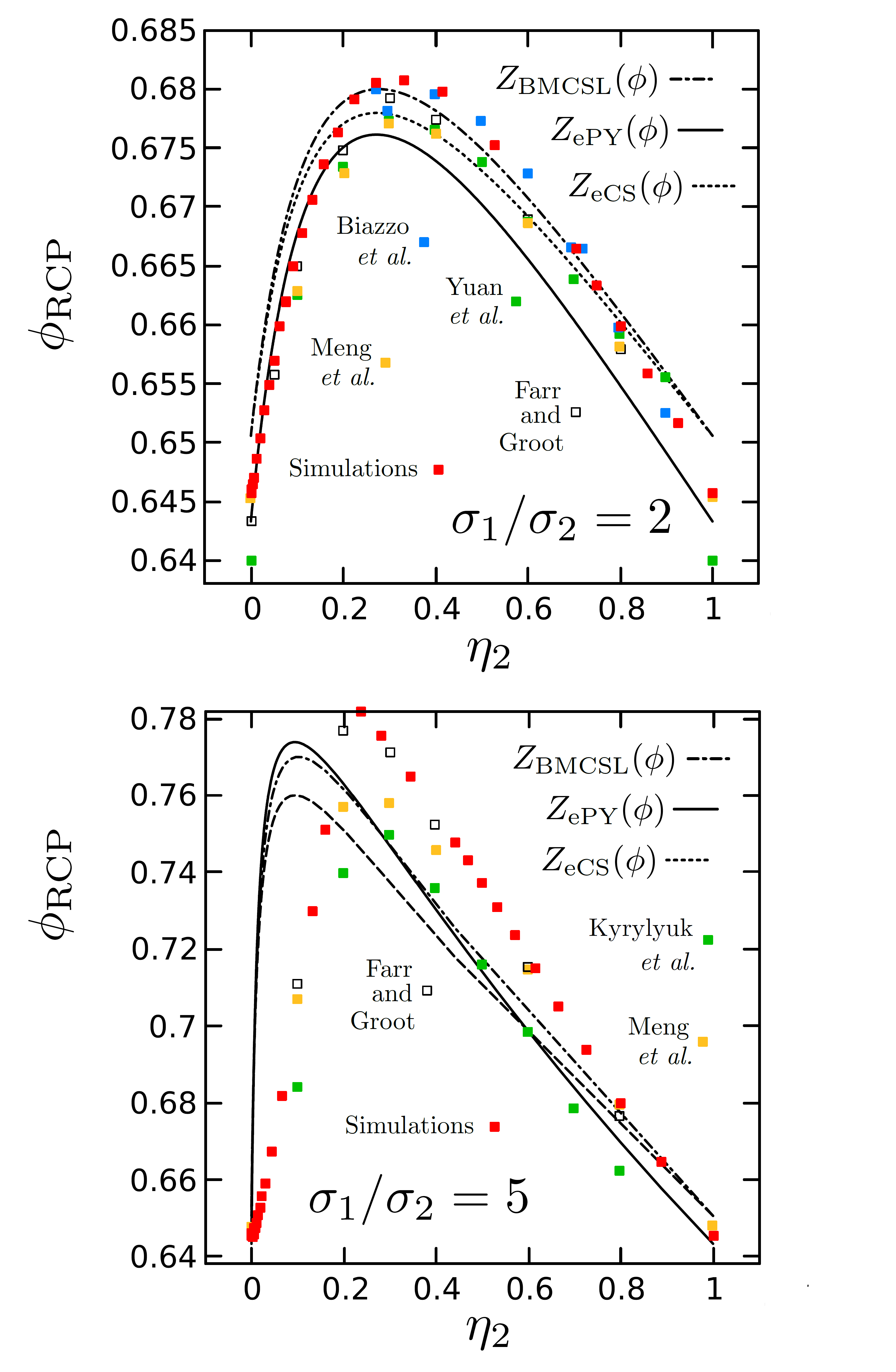}
	\caption{\textbf{RCP of a binary mixture of hard spheres.}
	Random close packing density $\phi_{\textrm{RCP}}$ against the volume fraction of spheres with diameter $\sigma_{2},$ in a binary mixture with fixed diameter ratio $\sigma_{1} / \sigma_{2} =2$ (top) and  $\sigma_{1} / \sigma_{2} =5$ (bottom).  Full, dashed and dot-dashed black lines represent results obtained when using the $Z_{\textrm{ePY}} (\phi),$ $Z_{\textrm{eCS}} (\phi)$ and $Z_{\textrm{BMCSL}} (\phi)$ approximations for the equation of state of the system, respectively, and the bcc configuration as a boundary condition to determine $C_{0}.$ Either in (top) and in (bottom) red points are simulations of our own, while yellow and white points are data adapted from Refs. \citep{MENG_BINARY_MIXTURE} and \citep{Farr_Groot_BINARY_MIXTURE}, respectively. Green points are adapted form Ref. \citep{Yuan_BINARY_MIXTURE} in (top) while are adapted from Ref. \citep{Kyrylyuk_BINARY_MIXTURE} in (bottom). Blue points in (top) are adapted from Ref. \citep{Biazzo}.} 
		\label{FIGURE_2_BINARY_MIXTURE}
\end{figure}

First, we focus on the shift $\Delta \phi_{\textrm{RCP}} \equiv \phi_{\textrm{RCP}} - \phi_{\textrm{RCP}}^{\textrm{mono}}$ induced by the discrete polydispersity on the RCP density of the pure sphere fluid, $\phi_{\textrm{RCP}}^{\textrm{mono}}$. In Fig. \ref{FIGURE_1_BINARY_MIXTURE}, we plot $\Delta \phi_{\textrm{RCP}}$ as a function of the volume fraction $\eta_2$ of small spheres, for several values of $\sigma_{1} / \sigma_{2} \in [1,2]$. 
We use solid, dashed and dot-dashed lines to represent results obtained when using the $Z_{\textrm{ePY}} (\phi),$ $Z_{\textrm{eCS}} (\phi)$ and $Z_{\textrm{BMCSL}} (\phi)$ approximations for the EOS $Z^{(m=2)} (\phi)$ of the system, respectively, with the bcc configuration used as a boundary condition to determine $C_0$.
Open points represent simulations from Ref.~\citep{Biazzo} while filled points are for our own simulations. 
For all the considered size ratios, our theory predicts the typical ``triangular" shape of the obtained density as a function of $\eta_{2}$.
Furthermore, a good match can be observed  between the numerical values and our predictions, with a disagreement of the same order of magnitude as the fluctuations between numerical sets of data.
These fluctuations, as well as the quantitative disagreement with our prediction, might have to do with the very small shifts in the RCP density which are hard to measure accurately using finite numbers of particles.
In fact, at size ratios very close to one, the differences between reported values for \textit{monodisperse} RCP are typically of the same order of magnitude as the shift due to polydispersity, which is why we here choose to plot the shift with respect to the monodisperse value.
For the case $\sigma_{1} / \sigma_{2} = 2,$ we show that our approach correctly captures the behavior of a binary granular system recently studied experimentally in Ref. \citep{Houfei_KOB} and represented by black star-shaped points in Fig. \ref{FIGURE_1_BINARY_MIXTURE}.

In Fig.~\ref{FIGURE_2_BINARY_MIXTURE} we now plot the \textit{absolute} (viz., not relative) value of $\phi_{\textrm{RCP}}$ as a function of $\eta_{2}$ in the cases $\sigma_{1} / \sigma_{2} =2 $ (top) and $\sigma_{1} / \sigma_{2} =5 $ (bottom).
For $\sigma_{1} / \sigma_{2} =2 $, we show good agreement with simulations over the whole range of volume fractions.
For $\sigma_{1} / \sigma_{2} =5 $, this time, we report agreement for $\eta_{2} >0.2$, but a rather strong deviation between our prediction and data for $\eta_{2} < 0.2$, where our prediction overestimates the packing fraction.
The cause of this disagreement is unclear, but it is worth mentioning that it is notoriously difficult to produce stable random packings in that region, as the system tends to form a jammed configuration of the large particles within which smaller particles can roam freely~\citep{Biazzo}.
A different choice of the EOS could also improve the agreement at large $\sigma_{1} / \sigma_{2}.$

Note that an EOS different from those used in this paper was recently considered as part of an analogous calculation in Ref. \citep{unoexplored_valley}.

\subsection{Continuous polydispersity\label{sec:continuous}}

Henceforth, we assume the particle diameter $\sigma$ to follow a continuous probability distribution $f (\sigma).$
We consider three different functional forms for $f(\sigma)$, which have been widely employed to describe polydispersity in colloidal systems \citep{groot,Hermes_2010,Berthier_distribution1,Berthier_distributionPRX}. 

We start by assuming the particle diameter $\sigma$ to follow the log-normal distribution \citep{cramer_BOOK}, for which results from numerical simulations are available in the literature \citep{groot,Hermes_2010}.
We use these numerical results to test our theoretical findings.
The log-normal distribution $f_{\textrm{log}} (\sigma)$ is defined as \citep{cramer_BOOK}
\begin{equation} \label{lognormal_f}
f_{\textrm{log}} (\sigma) = \frac{1}{\sigma \sqrt{2 \pi \alpha^{2}}} e^{ -  ( \ln  \sigma - \mu )^{2}/2 \alpha^{2} },
\end{equation} 
where $\alpha$ and $\mu$ are arbitrary parameters. The $n$-th moment $\left\langle \sigma^{n} \right\rangle  \equiv \int_{-\infty}^{\infty} d \sigma f_{\textrm{log}} (\sigma)  \sigma^{n}$ of $f_{\textrm{log}} (\sigma)$ is given by
\begin{equation} \label{lognorm_moments}
\left\langle \sigma^{n} \right\rangle = e^{ n \mu + n^{2} \alpha^{2} /2 },
\end{equation} 
such that the average value is $\left\langle \sigma \right\rangle = e^{ \mu + \alpha^{2} /2 }$ and the variance is $\textrm{var} [\sigma] \equiv \left\langle \sigma^{2} \right\rangle - \left\langle \sigma \right\rangle^{2} = e^{\alpha^{2} - 1 } e^{2 \mu + \alpha^{2}}$. The relative standard deviation can be written as
\begin{equation} \label{standard_dev_log}
s_{\sigma}^{\textrm{log}} \equiv \frac{\big( \left\langle \sigma^{2} \right\rangle  - \left\langle \sigma \right\rangle^{2} \big)^{1/2}}{\left\langle \sigma \right\rangle} =(  e^{\alpha^{2}} - 1 )^{1/2}.
\end{equation}
Insertion of Eq. \eqref{lognorm_moments} in the $Z_{\textrm{BMCSL}} (\phi),$ $Z_{\textrm{eCS}} (\phi)$ and $Z_{\textrm{ePY}} (\phi)$ approximations introduced in the previous section, yields three distinct approximate analytical expressions for the EOS $Z^{(m \to \infty)} (\phi)$ of our polydisperse system.
It can be easily verified the $Z^{(m \to \infty)} (\phi)$ thus obtained does not depend on the parameter $\mu$ but only on the parameter $\alpha.$
As it is clear from Eq. \eqref{standard_dev_log}, the relative standard deviation $s_{\sigma}^{\textrm{log}}$ of the $f_{\textrm{log}} (\sigma)$ distribution also depends exclusively on $\alpha$.
It follows that the effect of the polydispersity on the system can be fully described by either $\alpha$ or $s_{\sigma}^{\textrm{log}}$, for any arbitrary value of $\mu.$

\begin{figure}
    \centering
    \includegraphics[width=9.6cm]{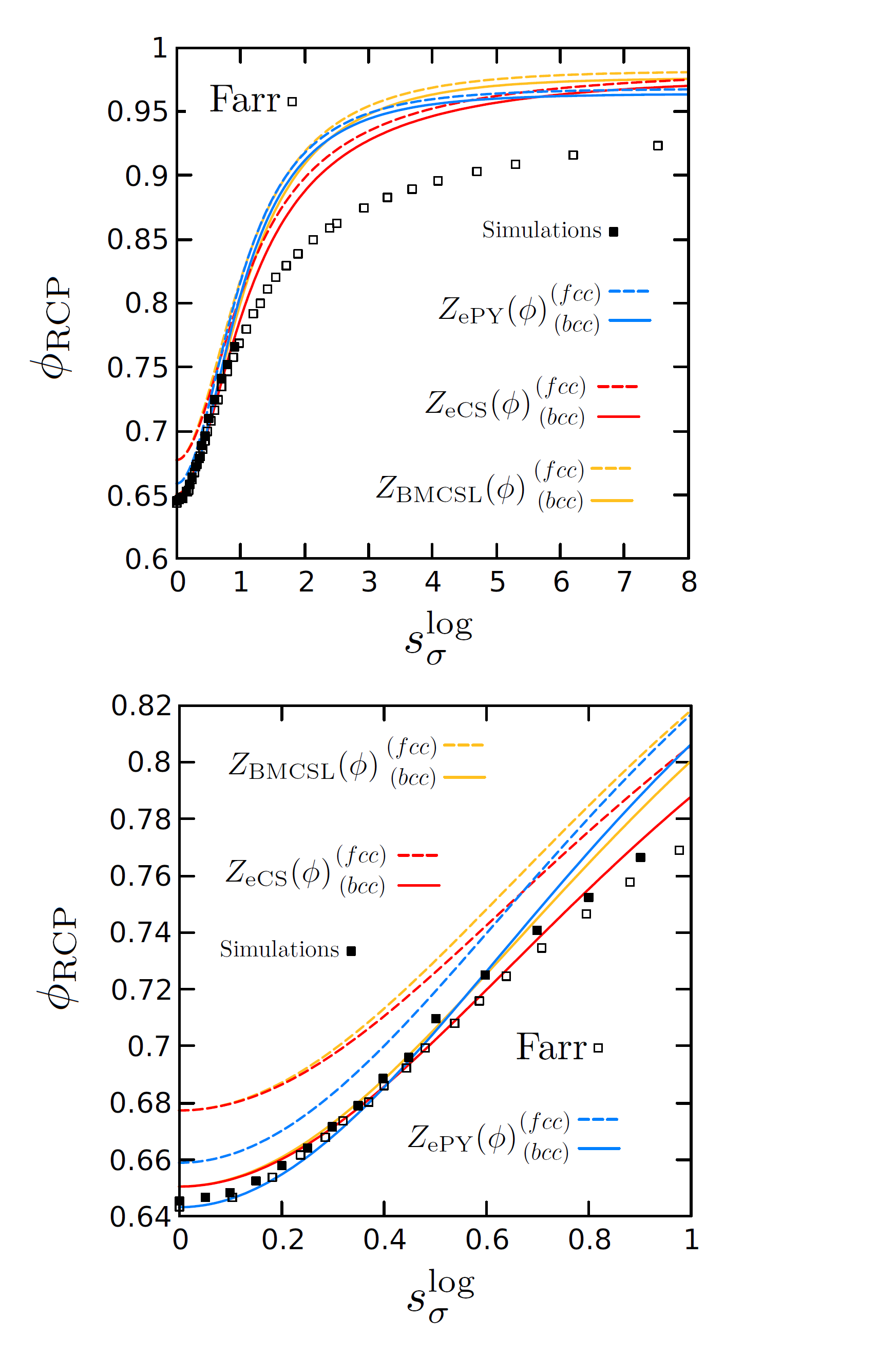}
        \caption{\textbf{RCP of log-normal-distributed hard spheres.} Random close packing density $\phi_{\textrm{RCP}}$ against the reduced standard deviation $s_\sigma^{\textrm{log}}$. Yellow, red and blue lines indicate results obtained when using the $Z_{\textrm{BMCSL}} (\phi), \ Z_{\textrm{eCS}} (\phi)$ and $Z_{\textrm{ePY}} (\phi)$ approximations for the equation of state of the system, respectively. Solid and dashed lines represent results obtained when using the bcc and fcc configurations, respectively, as a boundary condition to determine $C_0.$ White squares are data from Ref. \citep{groot}, while black squares are results from our own simulations. Top: Full range up to a final plateau. Bottom: Zoom on the small polydispersity region. %\sm{SM: the [10] in the plot is wrong}
  }
\label{fig:lognormalresults}
\end{figure}
In Fig.~\ref{fig:lognormalresults}, we show the predicted RCP density, $\phi_{\textrm{RCP}}$, against the reduced standard deviation $s_\sigma^{\textrm{log}}$.
We use yellow, red and blue lines to indicate the $\phi_{\textrm{RCP}}$ obtained using the $Z_{\textrm{BMCSL}} (\phi), \ Z_{\textrm{eCS}} (\phi)$ and $Z_{\textrm{ePY}} (\phi)$ approximations for the EOS  $Z^{(m \to \infty)} (\phi)$ of the system, respectively.
Moreover, we use solid and dashed lines to represent results obtained when the bcc and fcc configurations, respectively, are used as a boundary condition to determine $C_0 \equiv g_{0}/\sigma.$
We predict that $\phi_{\textrm{RCP}}$ increases monotonically with $s_\sigma^{\textrm{log}}$, until a plateau is reached.
Taking either of the proposed EOS, Eqs.~\eqref{ZBMCSL} or~\eqref{Zsantos}, in the limit of infinite skewness and variance predicts a limiting value of packing fraction, $\phi_{\textrm{max}}^{\textrm{log}} = 1/(1+C_0)\approx 0.97-0.98$, which reassuringly lies below the physical limit of $\phi=1$.
The increase of $\phi_{\textrm{RCP}}$ with the size polydispersity is in agreement with the fact that, when increasing polydispersity, smaller spheres typically fill the voids created between neighboring larger spheres, so that polydisperse hard-sphere fluids may reach larger packing fractions than monodisperse fluids \citep{Ogarko}.

These predictions are compared to both data from simulations adapted from Ref.~\cite{groot} (white squares), and to our own simulations (black squares). First, we note that a monotonic increase of $\phi_{\textrm{RCP}}$ as a function of $s_\sigma^{\textrm{log}}$ is also observed in simulations.
Furthermore, in the region $s_\sigma^{\textrm{log}} < 1$, we find good agreement between our predictions and results from both sets of simulations, as emphasized in the lower panel of Fig.~\ref{fig:lognormalresults}. Either choice of boundary condition (bcc or fcc) yield the right form as a function of $\Delta \phi_{\textrm{RCP}}$. The better agreement of the bcc curves can be attributed to the fact that the typical states found by the numerical compression protocols always lie below $\phi_{RCP}$ as defined in Fig.~\ref{fig:MRJLineSketch}, which is better approximated by the fcc curve, as shown in Fig.~\ref{fig:z_versus_phi}.

At larger polydispersities, there is growing disagreement between our predictions and numerical data.
We note that in the large polydispersity regime, it is very challenging to write a good approximate EOS, so that previous work typically designed piece-wise EOS to accommodate for large polydispersities~\cite{Santos2009}, and other choices of EOS than ours might work better at large $s_\sigma^{\textrm{log}}$. Furthermore, we note that it becomes increasingly challenging to obtain dense random jammed states as the polydispersity increases. 
This is illustrated, for instance, in Ref.~\cite{Baranau2014}, where slower and slower compression is required to approach the densest random packing as $s_\sigma$ increases.
Therefore, simulation results with finite compression rates always underestimate the actual maximal density, with an error that should become greater as the degree of polydispersity increases. In summary, both the EOS and the numerical results become progressively less reliable as $s_\sigma$ grows larger.

Having checked that our predictions hold for the log-normal distribution, we also consider in App.~\ref{app:GammaBerthier} two other common choices for $f (\sigma)$, namely a Gamma distribution and a truncated power-law distribution. We again find good agreement between the predicted and measured values for $s_\sigma < 1$.

\subsection{Universal behaviour at small polydispersity}

\begin{figure}
    \centering
    \includegraphics[width=1\columnwidth]{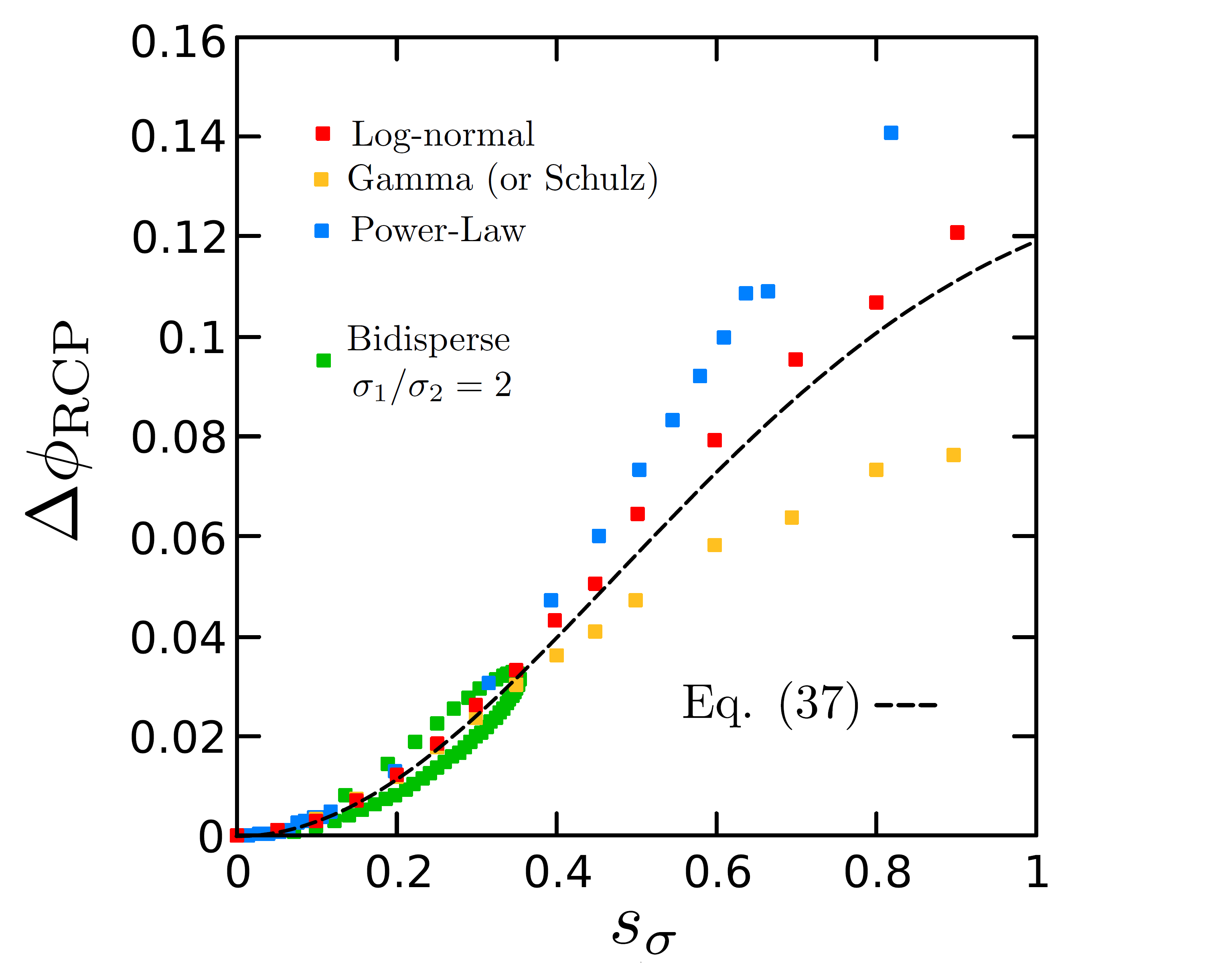}
    \caption{\textbf{Universal behaviour at small polydispersity.}
    With points we plot the numerically obtained shift $\Delta \phi_{\textrm{RCP}} \equiv \phi_{\textrm{RCP}} - \phi_{\textrm{RCP}}^{\textrm{mono}}$ at small polydispersity, for all four size distributions considered in this paper. The dashed line is the closed-form expression for the shift in $\phi_{\textrm{RCP}}$ for small polydispersity, Eq.~\eqref{eq:SmallPolySolution}, for the PY EOS and the bcc boundary condition.
    }
    \label{fig:smallpolyalldistris}
\end{figure}

It is worth noting that numerical data for all three continuous size distributions display remarkably similar shifts, $\Delta \phi_{\textrm{RCP}} \equiv \phi_{\textrm{RCP}} - \phi_{\textrm{RCP}}^{\textrm{mono}}$, in the limit of small polydispersity, $s_\sigma < 0.5$, as illustrated in Fig.~\ref{fig:smallpolyalldistris}.
In this figure, we also show data for a binary mixture, which forms a loop around the same universal trend.
This similarity suggests that the shift of RCP only depends on the second moment of the size distribution in the regime of small polydispersity, an effect which can be captured analytically from our approach.
Consider the approximate EOS used to construct the eCS and ePY expressions, Eq.~\eqref{Zsantos}.
In the limit of small polydispersity, $1 \gg \textrm{var}[\sigma]/\left\langle \sigma \right\rangle^2 \gg \textrm{skew}[\sigma] /\left\langle \sigma \right\rangle^3$, with $\textrm{skew}[\sigma] \equiv \left\langle \left(\sigma - \left\langle \sigma \right\rangle \right)^3 \right\rangle$ the skewness of the distribution.
We can approximate the EOS by taking its zero-skewness limit, and rewrite it as a function of $s_\sigma$
\begin{align}
    Z^{(m \to \infty)}(\phi,s_\sigma)&\approx Z(\phi) \frac{(1 + s_\sigma^2)(2 + 5 s_\sigma^2 + s_\sigma^4)}{2(1 + 3s_\sigma^2)^2} \nonumber \\
    &+ s_\sigma^2 \frac{5 + 12 s_\sigma^2 - s_\sigma^4 + 3 \phi (1 - s_\sigma^4)}{2 (1-\phi)(1 + 3s_\sigma^2)^2}.
\end{align}
This expression can be inserted into Eq.~\eqref{eq:SimplerCondition},
\begin{equation}
1 = C_0\left(Z^{(m \to \infty)}(\phi_{\textrm{RCP}},s_\sigma) - 1\right).
\label{eq:SmallPolyCondition}
\end{equation}
At small polydispersity, the packing fraction at RCP can be written as $\phi_{\textrm{RCP}} = \phi_{\textrm{RCP}}^{\textrm{mono}} + \Delta \phi_{\textrm{RCP}}$, with $\Delta\phi_{\textrm{RCP}} \ll 1$.
Taylor-expanding Eq.~\eqref{eq:SmallPolyCondition} to leading order in $\Delta \phi_{\textrm{RCP}}$ finally yields a closed-form small-polydispersity approximation
\begin{align}
    \Delta \phi_{\textrm{RCP}} \approx \frac{a_1 s_\sigma^2 + a_2 s_\sigma^4 + a_3 s_\sigma^6}{1 + b_1 s_\sigma^2 + b_2 s_\sigma^4 + b_3 s_\sigma^6},\label{eq:SmallPolySolution}
\end{align}
with coefficients that only depend on the monodisperse value of the RCP density, $C_0$ and the derivative of $Z$ at that density. The coefficients of this rational function are given in App.~\ref{app:Smalls}.

This approximation captures the universal parabolic dependence of the RCP density observed at small polydispersities in simulation data, as shown in Fig.~\ref{fig:smallpolyalldistris}.
In practice, this simplified expression could be useful in experimental contexts, in which the standard deviation of diameters is more easily accessible than the higher moments of the size distribution.

\begin{figure}
    \centering
    \includegraphics[width=1\columnwidth]{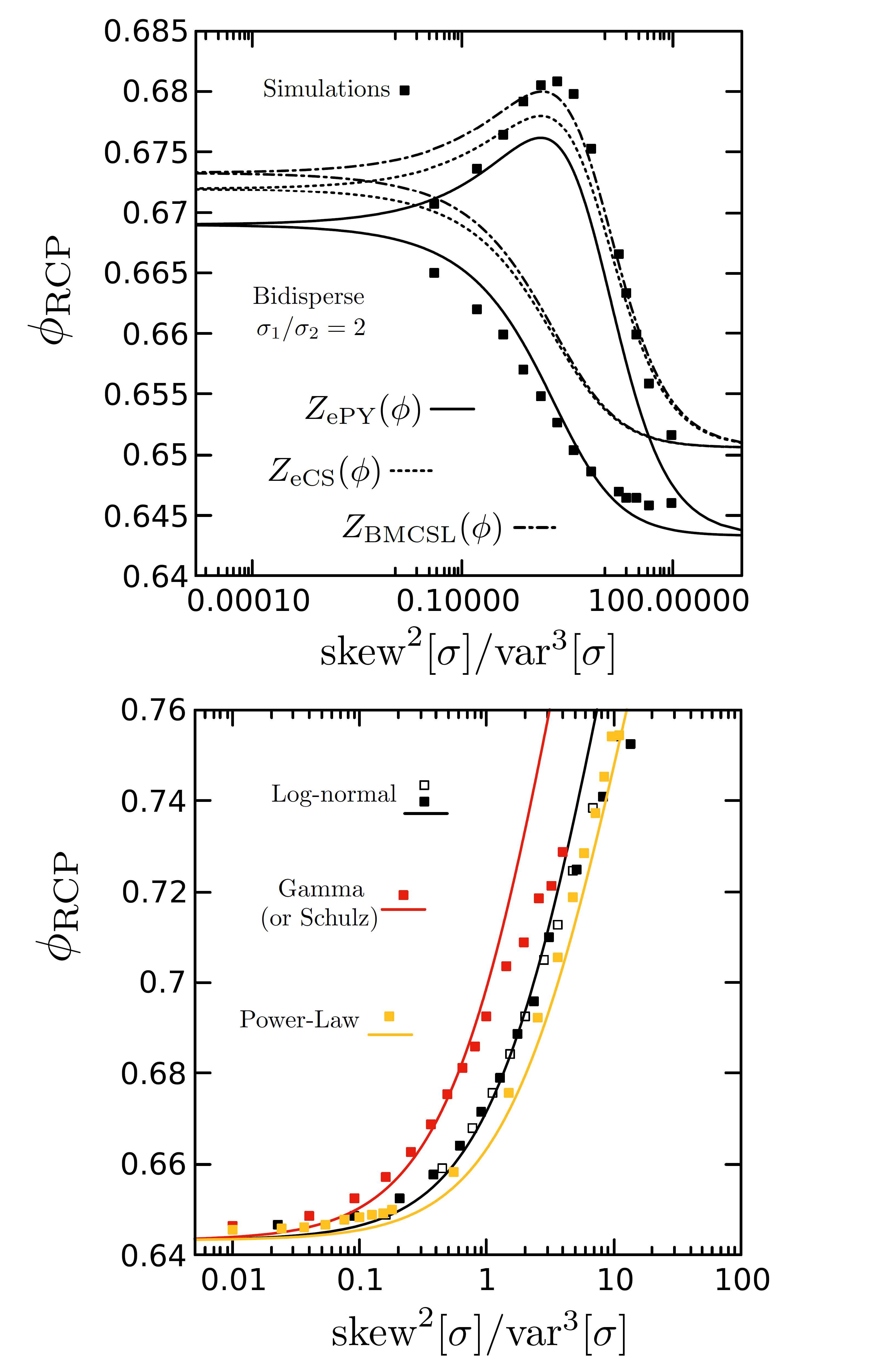}
    \caption{\textbf{Growth of the error with the skewness.}
    Log-linear plot of the numerically obtained $\phi_{\textrm{RCP}}$ for a bidisperse mixture (top) and all three considered continuous distributions (bottom) considered in this paper, against the dimensionless ratio of skewness to variance. Lines represent predictions from our theory for various equations of state (solid lines: ePY, dotted lines: eCS, dashed lines: BMCSL, all with a bcc boundary condition). In the bottom panel, white points are adapted from Ref. \citep{groot}.
    }
    \label{fig:skewness}
\end{figure}
Note that the closed-form expression, as well as the best agreement with data, is found in the limit of small skewness compared to the variance.
In Fig.~\ref{fig:skewness}, we check the validity of this statement for all tested distributions, showing that our predictions are best when $\text{skew}[\sigma]^{2}/\text{var}[\sigma]^{3}$ is small. Interestingly, this corresponds to intermediate number fractions of either species, or large variance, in the bidisperse case, but to small variance for the continuous distributions. These results highlight the importance of the choice of equations of state, which for polydisperse systems are generally designed for mixtures with small higher-order moments, as they are written as moment expansions~\cite{Santos2009}. Thus, it is possible that more faithful equations of state would lead to better results in the limit of large polydispersities.

\section{Conclusions and outlook  \label{sec:Conclusions_and_outlook}}

In this paper we investigated the effect of polydispersity on the random close packing (RCP) density $\phi_{\textrm{RCP}}$ of a hard-sphere fluid in $d=3$ dimensions.
The main insight of our approach is that we can arrive at a reasonable model of crowding for maximally random jammed states on the basis of approximate liquid theories.
This analogy is reminiscent of analogies between quenched disorder in type-II superconductors and thermal liquid structures~\cite{Sow1998, Mungan1998}, where a thermal average of the liquid theory matched the quenched average over disorder sufficiently well to get quantitative estimates of physical quantities.

This model of crowding allows us to estimate the effect of volume fraction on the contact value of the radial distribution function and, therefore, the kissing number, $z$. By combining this model for $z(\phi)$ with the isostaticity condition, $z=6$, required for the onset of shear rigidity at jamming \cite{Zaccone_ScossaRomano}, we derive a value for the RCP volume fraction $\phi_{\textrm{RCP}}$ for monodisperse hard-spheres. 

We show that a generalization of this approach to polydisperse systems amounts to a  straightforward substitution of the compressibility for a monodisperse hard sphere system, $Z(\phi)$, with its generalization to an $m$-component system, $Z^{(m)}(\phi)$, obtained from the generalization of an approximate equation of state to a mixture with a given choice of particle size distribution (either discrete or continuous). 

First, we consider a bidisperse distribution of particle sizes, and compare our predictions to data from a large selection of past works~\citep{Biazzo,MENG_BINARY_MIXTURE, Farr_Groot_BINARY_MIXTURE,Kyrylyuk_BINARY_MIXTURE,Yuan_BINARY_MIXTURE, Houfei_KOB}, as well as simulations of our own. For a wide range of size ratios and molar fractions, we observe good agreement between our theoretical predictions and the data. Then, we consider the particle diameter to follow one of three different types of continuous distribution widely used to approximate polydispersity in colloidal systems. In all cases, we find $\phi_{\textrm{RCP}}$ to increase monotonically with the relative standard deviation of the distribution, $s_{\sigma}$. We show that these predicted values are in good agreement with numerical results obtained from compression algorithms for polydispersities going up to $s_\sigma = 0.5$ (viz., $50\%$ standard deviation over mean ratio).
Moreover, we show that in the limit of small polydispersity, a closed-form expression for the RCP density that only depends on the reduced variance of the size distribution can be written, and accounts for universal behaviour observed for all tested size distributions.
We finally argue that the predictions become less reliable with increasing skewness over variance ratio, which is typically assumed to be small by the equations of states used in this paper. This raises the question of whether better equations of state for polydisperse systems could lead to better estimates.

More generally, this work raises an interesting numerical question worth investigating in future work: the precise determination of the location of the MRJ-line all the way to fcc, and the nature of states along it.
While states are routinely sampled either exactly at fcc, or on the isostatic line $z=6$ across densities~\cite{Atkinson2014,Ozawa2017}, it is extremely unlikely for usual compression schemes to end up anywhere between these two regimes, on the hyperstatic part of the MRJ-line.
One would therefore need to devise an algorithm to impose either minimal kissing numbers at a fixed density, or maximal density at a fixed kissing number.
Such work, while challenging, would shed light on the nature of the densest isostatic jammed packing, in particular on its fundamental ties with glassiness~\cite{Parisi_Zamponi, Baranau2014} and critical points of absorbing-state models~\cite{Hexner2018,Wilken2021}.

% It is interesting to highlight possible ties between the definition of $\phi_{\textrm{RCP}}$ presented in this paper, and the density of the so-called Glass Close-Packing (GCP).
% The latter is tied to the \textit{ideal} glassy state of hard spheres.
% Within the context of replica theory \citep{Parisi_Zamponi,Parisi_Zamponi_JCP}, such a state has been conjectured to stabilize into a hard-sphere fluid as a result of an ideal glass transition occurring in the metastable continuation of the liquid branch above $\phi_{\textrm{freeze}}.$
% In contrast to the \textit{dynamic} glass transition, which ``merely" signals the point at which the structural relaxation time becomes comparable to experimental time scales, the ideal glass transition is a true thermodynamic transition to an equilibrium state of matter \citep{CAVAGNA200951}.
% It is often stated that GCP, the jammed endpoint of the ideal glass branch, is the end of the so-called ``J-line'' of glassy jammed states, which is typically isostatic all along~\cite{Baranau2014,Ozawa2017,Charbonneau2017}.
% Our definition of RCP, as the densest isostatic jammed packing, may coincide with GCP, provided that no non-glassy isostatic jammed states exists at larger densities.
% %If so, the analogy presented in this paper allows to study the evolution of the GCP density with polidispersity directly in a finite dimension of space.

Finally, the introduced theoretical scheme could be used to investigate the \textit{additional} jamming line recently found for binary mixtures of hard spheres in Refs. \citep{Sperl_additional_jamming, Ikeda_additional_jamming}. Furthermore, using known equations of states, it could be applied not only to arbitrary polydispersity of hard spheres, but also to other particle shapes, which could serve as a simple tool to understand the jamming transition of general hard objects.  

\begin{acknowledgments}
The authors thank Daan Frenkel for useful discussions in the preliminary stages of this work, as well as David Grier for interesting suggestions. C.A. gratefully acknowledges financial support from Syngenta AG. A.Z. gratefully acknowledges funding from the European Union through Horizon Europe ERC Grant number: 101043968 ``Multimech'', and from US Army Research Office through contract nr.   W911NF-22-2-0256. M.C. and S.M. acknowledge the Simons Center for Computational Physical Chemistry for financial support.
This work was supported in part through the NYU IT High Performance Computing resources, services, and staff expertise.
S.M. was partially supported by National Science Foundation grant IIS-2226387, and performed part of this work at the Aspen Center for Physics, which is supported by National Science Foundation grant PHY-1607611.
\end{acknowledgments}

\section*{Data availability}
The data that support the findings of this study are available from the corresponding authors upon reasonable request.

\appendix

\section{Equations of state\label{app:EoS}}

We here list the equations of state used in the main text within our analogy between jammed states and equilibrium configurations of hard spheres.

\subsection{Monodisperse equations of state}

For a monodisperse hard-sphere system in three dimensions, from the analytical solution of the Percus-Yevick (PY) equation for the direct correlation function, two analytical EOS can be obtained \citep{Hansen_McDonald_BOOK}.
By injecting the PY solution into the compressibility equation, the \textit{compressibility} EOS, $Z_{\textrm{PY}}^{c} (\phi)$, is derived, while by injecting it into the virial expansion the \textit{virial} EOS, $Z_{\textrm{PY}}^{v} (\phi)$, is obtained.
Thiele \citep{thiele} and Wertheim \citep{wertheim} independently found the compressibility and the virial equations of state to be given by $ Z^{c}_{\textrm{PY}} (\phi) = (1 + \phi + \phi^{2}) / (1- \phi)^{3}$ and $Z^{v}_{\textrm{PY}} (\phi) = (1 + 2 \phi + 3 \phi^{2}) / (1- \phi)^{2},$ respectively.
Subsequently, Carnahan and Starling \citep{CS} showed that a more accurate EOS for hard spheres is given by a linear combination of $Z^{v}_{\textrm{PY}}  (\phi)$ and $Z^{c}_{\textrm{PY}}  (\phi),$ and introduced the so-called Carnahan-Starling (CS) EOS, $Z_{\textrm{CS}} (\phi) \equiv \frac{2}{3} Z^{c}_{\textrm{PY}}  (\phi) + \frac{1}{3} Z^{v}_{\textrm{PY}} (\phi) = (1+\phi +\phi^{2} - \phi^{3})/(1-\phi)^{3}.$
Upon insertion of the $Z^{v}_{\textrm{PY}} (\phi)$ EOS into Eq. \eqref{Z_mono}, one obtains \citep{Song}
\begin{equation} \label{PY}
g_{\textrm{PY}} (\sigma; \phi) = \frac{1 + \phi/2}{(1 - \phi)^{2}},
\end{equation} 
while insertion of the $Z_{\textrm{CS}} (\phi)$ EOS into Eq. \eqref{Z_mono}, leads to
\begin{equation} \label{CS}
g_{\textrm{CS}} (\sigma; \phi) = \frac{1 -  \phi/2}{(1 - \phi)^{3}}.
\end{equation}  
Likewise, phenomenological equations of state with numerical fitting factors have been proposed to match numerical data on the equilibrium fcc branch of hard spheres, that diverges at fcc.
For instance the Young and Alder (YA) equation of state reads~\cite{Young1979a,mulero},
\begin{equation}
Z_{YA}(\alpha) = \frac{3}{\alpha} + 2.81 + 0.47 \alpha - 1.36 \alpha^2 + 6.41 \alpha^3,
\end{equation}
with $\alpha = (\phi_{\textrm{fcc}} - \phi)/\phi$.

\subsection{Polydisperse equations of state}

In this paper, we consider three different equations of state for mixtures of hard spheres at equilibrium.
The first one is the Boubl\'ik-Mansoori-Carnahan-Starling-Leland (BMCSL) EOS, which reads \citep{boublik,MCSL}
\begin{equation} \label{ZBMCSL}
\begin{aligned}
Z_{\textrm{BMCSL}} (\phi) &= \frac{1}{1-\phi} + \frac{3 \phi}{(1-\phi)^{2}} \frac{\left\langle \sigma \right\rangle \left\langle \sigma^{2} \right\rangle}{\left\langle \sigma^{3} \right\rangle}  \\
&+ \frac{\phi^{2} (3 - \phi)}{(1-\phi)^{3}} \frac{\left\langle \sigma^{2} \right\rangle^{3}}{\left\langle \sigma^{3} \right\rangle^{2}},
\end{aligned}
\end{equation}
and reduces to the CS EOS $Z_{\textrm{CS}} (\phi),$ in the monodisperse limit $f (\sigma) = \sum_{i=1}^{m} x_{i} \delta (\sigma_{i} - \sigma)$ with $m =1.$
To get two other candidates for the EOS, we follow the recipe introduced by Santos et al. in Ref. \citep{santos} to derive the EOS $Z^{(m \to \infty)} (\phi)$ of a polydisperse mixture of additive hard spheres in terms of the EOS $Z (\phi)$ of a one-component system,
\begin{equation} \label{Zsantos}
\begin{aligned}
Z^{(m 	\to \infty)} & (\phi) = 1 + \big[ Z (\phi) - 1 \big] \frac{\left\langle \sigma^{2} \right\rangle}{2 \left\langle \sigma^{3} \right\rangle^{2}} \big( \left\langle \sigma^{2} \right\rangle^{2} + \left\langle \sigma \right\rangle \left\langle \sigma^{3} \right\rangle \big) \\
&+ \frac{\phi}{(1-\phi)} \bigg[ 1 - \frac{\left\langle \sigma^{2} \right\rangle}{\left\langle \sigma^{3} \right\rangle^{2}}  \big( 2 \left\langle \sigma^{2} \right\rangle^{2} - \left\langle \sigma \right\rangle \left\langle \sigma^{3} \right\rangle \big) \bigg].
\end{aligned}
\end{equation}

In this paper we consider the cases $Z (\phi)=Z_{\textrm{CS}} (\phi)$ and $Z (\phi)=Z_{\textrm{PY}}^{v} (\phi)$, which respectively yield the so-called extended Carnahan-Starling (eCS) EOS,
\begin{equation} \label{eCS} 
\begin{aligned}
Z_{\textrm{eCS}} (\phi) &= Z_{\textrm{BMCSL}} (\phi) + \frac{\phi^{3}}{(1-\phi)^{3}} \frac{\left\langle \sigma^{2} \right\rangle}{\left\langle \sigma^{3} \right\rangle^{2}} \big( \left\langle \sigma \right\rangle \left\langle \sigma^{3} \right\rangle - \left\langle \sigma^{2} \right\rangle^{2} \big),
\end{aligned} 
\end{equation}
and extended Percus-Yevick (ePY) EOS, $Z_{\textrm{ePY}} (\phi)$.
% \begin{align}
% &Z_{\textrm{ePY}} (\phi) =  Z_{\textrm{BMCSL}} (\phi) \nonumber \\ 
% &+ \frac{\phi}{\left(1 - \phi\right)^3} \frac{\left\langle \sigma^2\right\rangle}{\left\langle \sigma^3 \right\rangle^2} \left( \left\langle \sigma^2\right\rangle^2 \left(\phi^2 - 6 \phi + 3 \right) 
% - 2 \left\langle \sigma \right\rangle \left\langle \sigma^{3} \right\rangle \left(1 - \phi\right)^2 \right).
%      \label{ePY}
% \end{align}
By construction, $Z_{\textrm{eCS}} (\phi)$ and $Z_{\textrm{ePY}} (\phi)$ reduce to the  $Z_{\textrm{CS}} (\phi)$ and $Z_{\textrm{PY}} (\phi)$ EOS, respectively, in the monodisperse limit.

\section{Gamma and Truncated Power-law distributions \label{app:GammaBerthier}}

In the main text, we present a full set of results for continuous polydispersity drawn from the log-normal distribution, then briefly discuss results for two other common distributions.
In this appendix, we show the full set of results for these distributions.

\begin{figure}
  \centering
  \includegraphics[width=8.6cm]{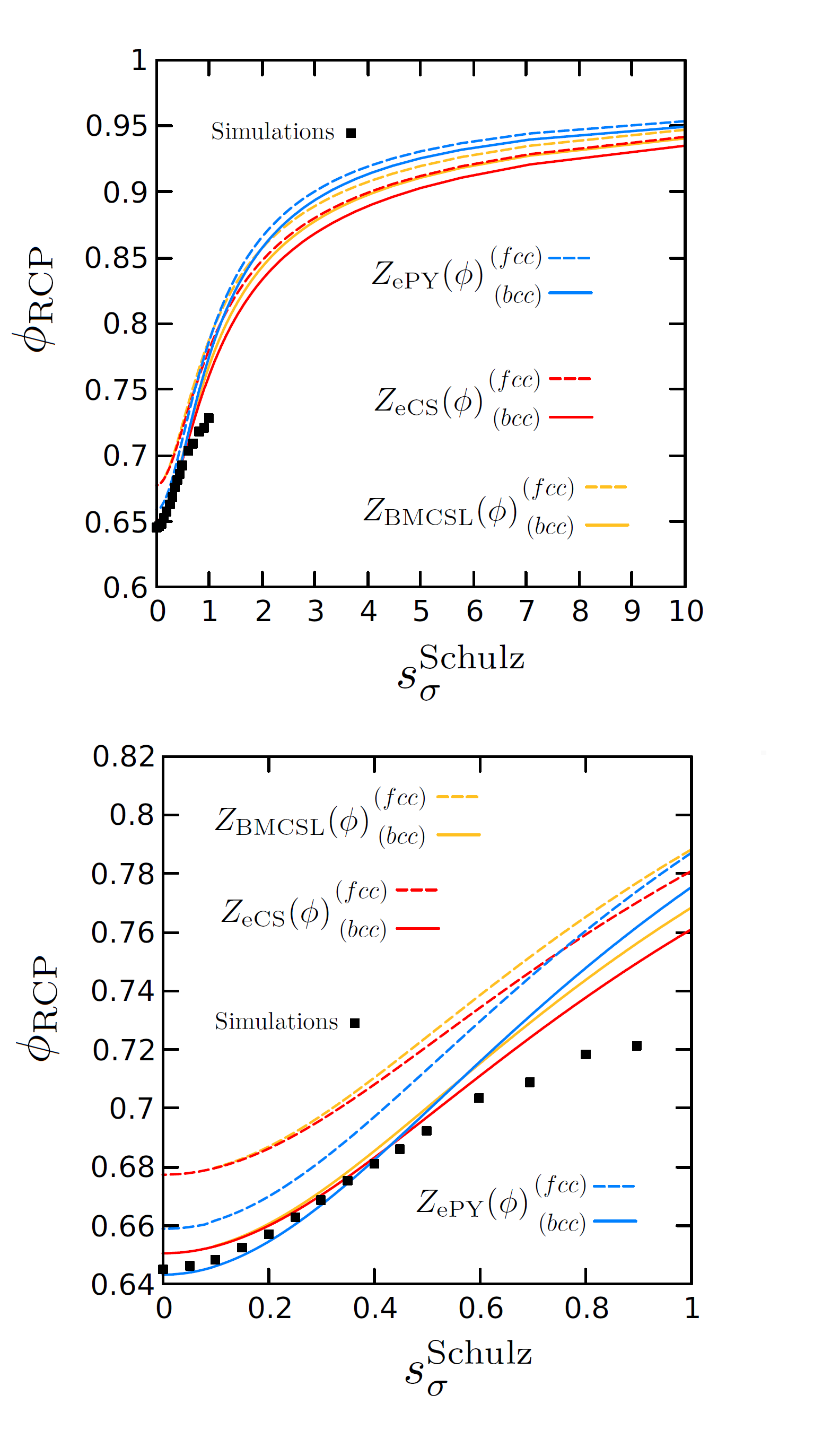}
  \caption{\textbf{RCP of gamma-distributed hard spheres.}
  Random close packing density $\phi_{\textrm{RCP}}$ against the reduced standard deviation $s_\sigma^{\textrm{Schulz}}$ for hard spheres with diameter following the Gamma, or Schulz distribution. Top: full range up to a final plateau. Bottom: Zoom on the small polydispersity region.
  }
 \label{fig:Gamma}
\end{figure}
\begin{figure}
  \centering
  \includegraphics[width=9cm]{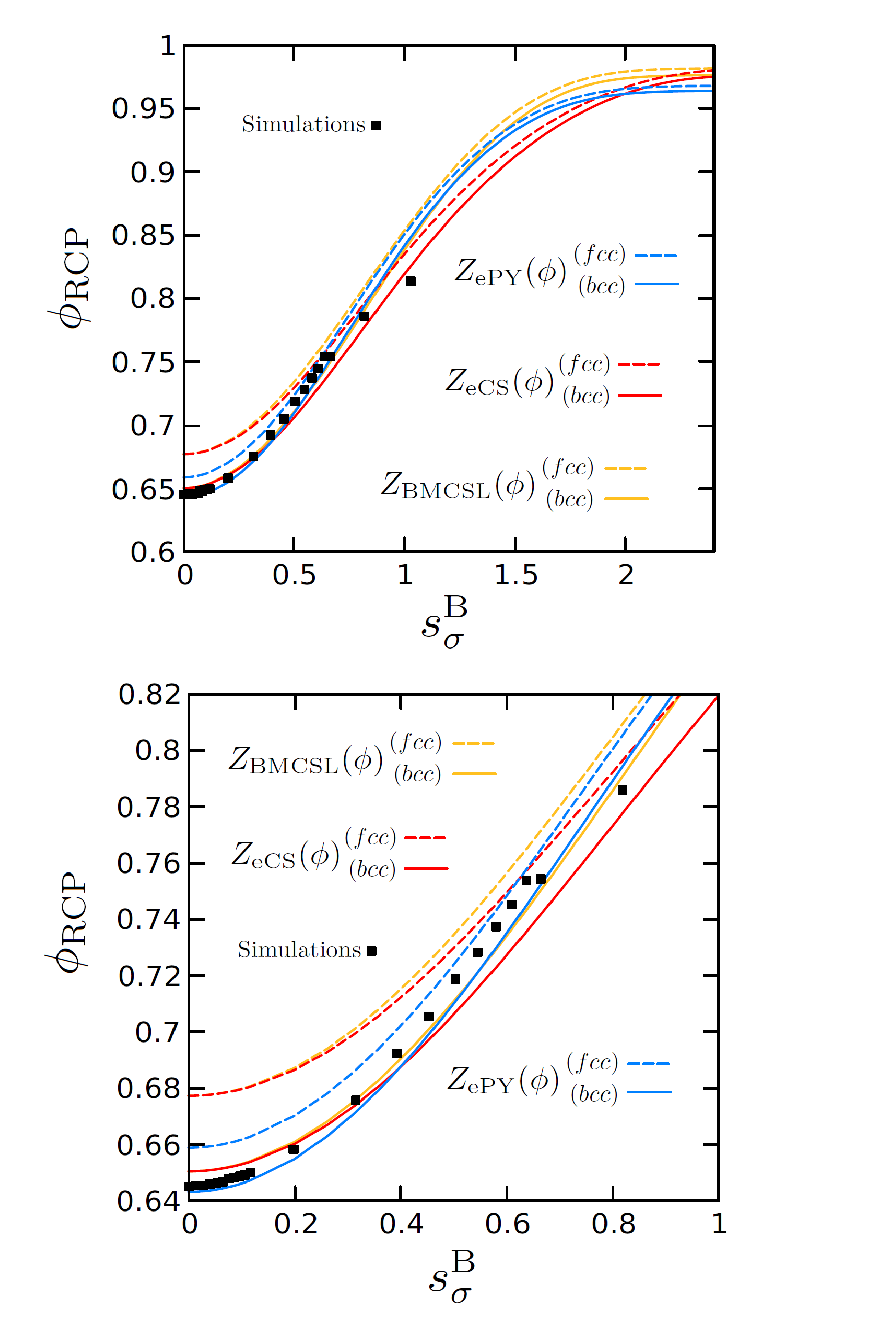}
  \caption{\textbf{RCP of power-law-distributed hard spheres.}
  Random close packing density $\phi_{\textrm{RCP}}$ against the reduced standard deviation $s_\sigma^{B}$ for hard spheres with diameter following a truncated power-law. Top: full range up to a final plateau. Bottom: Zoom on the small polydispersity region.
  }
 \label{fig:GammaBerthier}
\end{figure}

The first one is the Gamma distribution, also called the Schulz distribution in this context \cite{Schultz}, which reads
\begin{equation} \label{schulz_distribution}
f_{\textrm{Schulz}} (\sigma) = \bigg( \frac{\alpha +1}{\bar{\sigma}} \bigg)^{\alpha + 1} \frac{\sigma^{\alpha} e^{- (\alpha +1 ) \sigma / \bar{\sigma}}}{\Gamma (\alpha +1)},
\end{equation}
where $\Gamma (z)$ is the gamma function \citep{Abramovitz_BOOK}. The moments of $f_{\textrm{Schulz}} (\sigma)$ are given by
\begin{equation} \label{Schulz_moments}
\left\langle \sigma^{j} \right\rangle = \frac{\Gamma (j+\alpha+1)}{\Gamma (\alpha +1)} \bigg( \frac{\bar{\sigma}}{\alpha +1} \bigg)^{j},
\end{equation}
such that the average value is $\left\langle \sigma \right\rangle = \bar{\sigma}$ and the variance is $\textrm{var} [\sigma] \equiv \left\langle \sigma^{2} \right\rangle - \left\langle \sigma \right\rangle^{2} = \bar{\sigma}^{2} / (1+ \alpha)$. The relative standard deviation can be written as
\begin{equation}
s_{\sigma}^{\textrm{S}} = \frac{\big( \left\langle \sigma^{2} \right\rangle  - \left\langle \sigma \right\rangle^{2} \big)^{1/2}}{\left\langle \sigma \right\rangle} = \frac{1}{\big( \alpha +1 \big)^{1/2}}.
\end{equation}

The last distribution we consider is a truncated power-law distribution, which scales as the inverse of the occupied volume, introduced by Berthier and co-workers in the context of supercooled liquids \citep{Berthier_distribution1,Berthier_distributionPRX},
\begin{equation} \label{berthier_distribution}
f_{\textrm{B}} (\sigma) = \frac{A}{\sigma^{3}},
\end{equation}
where $A$ is a normalizing constant and $\sigma \in [\sigma_{\textrm{min}} , \sigma_{\textrm{max}}]$ with $\sigma_{\textrm{min}}$ and $\sigma_{\textrm{max}}$ the minimum and maximum diameter values, respectively.
By imposing the normalization condition $\int_{\sigma_{\textrm{min}}}^{\sigma_{\textrm{max}}} f_{\textrm{B}} (\sigma) =1,$ it follows that $A = 2 \sigma_{\textrm{min}}^{2} \sigma_{\textrm{max}}^{2} / ( \sigma_{\textrm{max}}^{2} - \sigma_{\textrm{min}}^{2} ).$
The mean value and the variance of the distribution are $\left\langle \sigma \right\rangle = 2 \sigma_{\textrm{min}} \sigma_{\textrm{max}} / ( \sigma_{\textrm{min}} +  \sigma_{\textrm{max}} ) $ and $\textrm{var} [\sigma] = - 4 \sigma_{\textrm{min}}^{2} \sigma_{\textrm{max}}^{2} / ( \sigma_{\textrm{min}} + \sigma_{\textrm{max}} )^{2} + 2 \sigma_{\textrm{min}}^{2} \sigma_{\textrm{max}}^{2} \ln ( \sigma_{\textrm{min}} / \sigma_{\textrm{max}} ) / ( \sigma_{\textrm{min}}^{2} - \sigma_{\textrm{max}}^{2} ) ,$ respectively. By introducing $\beta \equiv \sigma_{\textrm{max}}/ \sigma_{\textrm{min}},$ the relative standard deviation can be written as
\begin{equation}
s_{\sigma}^{\textrm{B}} = \frac{\big( \left\langle \sigma^{2} \right\rangle  - \left\langle \sigma \right\rangle^{2} \big)^{1/2}}{\left\langle \sigma \right\rangle} = \bigg( \frac{1+\beta}{2 (\beta-1)} \ln \beta -1 \bigg)^{1/2}.
\end{equation}

Like in the case of the log-normal distribution, we take advantage of the explicit knowledge of the moments of the $f_{\textrm{Schulz}} (\sigma)$ and the $f_{\textrm{B}} (\sigma)$ distributions to compute the EOS of the system, for each of the approximations considered in the previous section.
We observe that again $Z^{(m\to \infty)} (\phi)$ only depends on a single parameter representing the spread of the distribution.
This is the $\alpha$ parameter in the Schulz distribution \eqref{schulz_distribution} and $\beta$ in the distribution of Berthier and co-workers \eqref{berthier_distribution}.
We then follow the same protocol of the log-normal distribution to find $\phi_{\textrm{RCP}}$ as a function of the size polydispersity, expressed in terms of the reduced standard deviation.

The results for the Gamma and the truncated power-law distributions are shown in Figs. \ref{fig:Gamma} and \ref{fig:GammaBerthier}, respectively.
The same color and line-style codes as in Fig.~\ref{fig:lognormalresults} are used therein to show predictions of RCP using different EOS and boundary conditions.
We find results qualitatively similar to those discussed in the case of the log-normal distribution, with quantitative differences in both the rate of increase of the RCP packing fraction, and the precise value of the large-polydispersity plateau.
Furthermore, we show values of $\phi_{\textrm{RCP}}$ measured from our own simulations as symbols.

\section{Analytical expression at small polydispersity \label{app:Smalls}}

In the main text, we present an explicit analytical expression for the shift of the RCP density at small polydispersity.
We here give its complete expression,
\begin{align}
    \Delta \phi_{\textrm{RCP}} \approx \frac{a_1 s_\sigma^2 + a_2 s_\sigma^4 + a_3 s_\sigma^6}{1 + b_1 s_\sigma^2 + b_2 s_\sigma^4 + b_3 s_\sigma^6},\label{eq:SmallPolySolutionApp}
\end{align}
with 
\begin{align}
    a_1 &= \frac{5 -  \phi_{\textrm{RCP}}^{\textrm{mono}} (5 + 8 C_0)}{2C_0 (1 - \phi_{\textrm{RCP}}^{\textrm{mono}})Z'(\phi_{\textrm{RCP}}^{\textrm{mono}})}, \\
    a_2 &= 6 \frac{1 - (C_0 + 1)\phi_{\textrm{RCP}}^{\textrm{mono}}  }{C_0 (1 - \phi_{\textrm{RCP}}^{\textrm{mono}}) Z'(\phi_{\textrm{RCP}}^{\textrm{mono}}) }, \\
    a_3 &= \frac{(4 C_0 + 1) \phi_{\textrm{RCP}}^{\textrm{mono}} - 1}{2C_0 (1 - \phi_{\textrm{RCP}}^{\textrm{mono}})Z'(\phi_{\textrm{RCP}}^{\textrm{mono}}) }, \\
    b_1 &= \frac{7}{2} + \frac{4}{(1 - \phi_{\textrm{RCP}}^{\textrm{mono}})^2 Z'(\phi_{\textrm{RCP}}^{\textrm{mono}})}, \\
    b_2 &= 3 +  \frac{6}{(1 - \phi_{\textrm{RCP}}^{\textrm{mono}})^2 Z'(\phi_{\textrm{RCP}}^{\textrm{mono}})}, \\
    b_3 &= \frac{1}{2} -  \frac{2}{(1 - \phi_{\textrm{RCP}}^{\textrm{mono}})^2 Z'(\phi_{\textrm{RCP}}^{\textrm{mono}})},
\end{align}
where the derivative of the EOS for instance takes the values
\begin{align}
    {Z_{\textrm{CS}}}'(\phi_{\textrm{RCP}}^{\textrm{mono}}) &= \frac{4 + 2\phi_{\textrm{RCP}}^{\textrm{mono}} (2 - \phi_{\textrm{RCP}}^{\textrm{mono}}) }{(1 - \phi_{\textrm{RCP}}^{\textrm{mono}})^4}, \\
    {Z^{c}_{\textrm{PY}}}'(\phi_{\textrm{RCP}}^{\textrm{mono}}) &= \frac{4 + 8 \phi_{\textrm{RCP}}^{\textrm{mono}}}{(1 - \phi_{\textrm{RCP}}^{\textrm{mono}})^3},
\end{align}
using the CS and compressibility PY EOS, respectively.

\section{Numerical methods\label{app:numerics}}

\begin{figure}
    \centering
    \includegraphics[width=0.96\columnwidth]{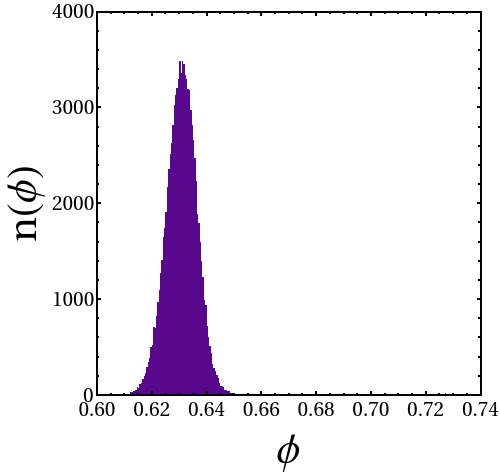}
    \caption{\textbf{Histogram of output densities of the TJ algorithm.}
    This histogram was obtained by performing about $10^5$ compressions of $N=108$ hard spheres.}
    \label{fig:densityhistogram}
\end{figure}

We here describe the numerical simulations used to validate our predictions of $\phi_{\textrm{RCP}}$.
In the qualitative validation of the analogy with equilibrium liquid, the data used in Fig.~\ref{fig:z_versus_phi} was generated using a Torquato-Jiao (TJ) algorithm~\cite{Jiao2011,Atkinson2014}.
This algorithm starts from a low-density isotropic state, in our case, following Ref.~\cite{Atkinson2013}, a $\phi = 0.1$ arrangement of monodisperse spheres generated by random sequential adsorption (RSA).
It then proposes isotropic compression, simple shear, and particle motion in such a way that density gain is optimized at every step, with the constraint that each type of move has an amplitude bounded by a user-defined value.
The direction of motion is determined by a user-defined interaction radius around particles, a so-called ``sphere of influence"~\cite{Jiao2011}.
At each step, particles look up neighbors that lie within that sphere, then move towards their center of mass (or, in the case of a single neighbor, the move is performed away from the particle).
In our case, we set maximal compression, shear, and displacement amplitudes to $0.01$ times the diameter of a particle, and we make the radius of the sphere of influence $3.5$ particle diameters.
These values were set using Refs.~\cite{Jiao2011,Atkinson2013} to favor higher densities.
The program ends when the volume change between two steps changes by less than $2\times10^{-12}$ in units of diameters cubed.
As mentioned in the main text, this algorithm favors maximally random configurations, but also overwhelmingly generates densities around a central one, at roughly $0.63-0.64$, see the histogram in Fig.~\ref{fig:densityhistogram}.
That is why we choose a relatively small number of particles $N=108$, so that we can get a large set of independent compression events and manage to measure final states far away from the mean of that histogram.

To generate the data in the polydisperse case, we use a variation of the Lubachevsky-Stillinger (LS) algorithm~\cite{Lubachevsky1990, Lubachevsky1991}, introduced in Ref.~\cite{Baranau2014}, in which dense random packings are obtained using increasingly slow compression, alternated with free evolution to let the pressure of the system relax to smaller values every time it crosses the threshold value $10^{12}$.
In practice, we used the same code and followed the same recipe as in Ref.~\cite{Baranau2014}: starting from random positions obtained by Poisson point-picking in a cubic box, we pre-compressed particles to a target packing fraction of $0.4-0.6$ using a force-biased algorithm.
We then ran a first, fixed-rate LS algorithm, at compression rate $\gamma$.
We finally ran the modified LS algorithm (MLS), yielding a final packing fraction $\phi_{\textrm{MLS}}(\gamma)$ that depends on the compression rate of the preliminary fixed-rate compression.
The RCP packing fractions presented in the text are values of the density estimated from an extrapolation of the observed trend $\phi_{\textrm{MLS}}(\gamma)$ in the limit $\gamma\to 0$.
In our simulations, we used $N = 10^4$ particles and, using Fig. 2 of Ref.~\cite{Baranau2014} as a guide, we used inverse compression rates in the range $\gamma^{-1} \in \left[10^2; 10^5\right]$ in the LS algorithm, except for $s_\sigma < 0.1$ where we used a maximal inverse rate of $2 \times 10^3$ to avoid crystallization.

%\nocite{*}
\bibliography{references} %, Bibtex-PolyRCP_ExtraRefs}% Produces the bibliography via BibTeX.

\clearpage

\end{document}